\documentclass[12pt,preprint]{aastex}
\usepackage{amsmath}
\usepackage{natbib}
\usepackage{hyperref}
\usepackage{cleveref}
\usepackage{rotating}
\usepackage{color}
\usepackage{enumerate}

\begin{document} 
\bibliographystyle{plainnat}

\title{\Large{Gamma-ray burst prompt correlations: selection and instrumental effects}}

\author{Dainotti M. G.\altaffilmark{1,2,3}}

\altaffiltext{1}{Physics Department, Stanford University, Via Pueblo Mall 382, Stanford, CA, USA, E-mail: mdainott@stanford.edu}
\altaffiltext{2}{INAF-Istituto di Astrofisica Spaziale e Fisica cosmica, Via Gobetti 101, 40129,Bologna, Italy}
\altaffiltext{3}{Astronomical Observatory, Jagiellonian University, ul. Orla 171, 31-244 Krak{\'o}w, Poland. E-mails: mariagiovannadainotti@yahoo.it}

\begin{abstract}

The prompt emission mechanism of gamma-ray bursts (GRB) even after several decades remains a mystery. However, it is believed that correlations between observable GRB properties given their huge luminosity/radiated energy and redshift distribution extending up to at least $z \approx 9$, are promising possible cosmological tools. They also may help to discriminate among the most plausible theoretical models. Nowadays, the objective is to make GRBs standard candles, similar to supernovae (SNe) Ia, through well-established and robust correlations. However, differently from SNe Ia, GRBs span over several order of magnitude in their energetics and hence cannot yet be considered standard candles. Additionally, being observed at very large distances their physical properties are affected by selection biases, the so called Malmquist bias or Eddington effect. We describe the state of the art on how GRB prompt correlations are corrected for these selection biases in order to employ them as redshift estimators and cosmological tools. We stress that only after an appropriate evaluation and correction for these effects, the GRB correlations can be used to discriminate among the theoretical models of prompt emission, to estimate the cosmological parameters and to serve as distance indicators via redshift estimation.
\end{abstract}

\maketitle
\pagebreak
\tableofcontents

\pagebreak

\section{Introduction}
Gamma-ray bursts (GRBs), discovered in late 1960's \citep{klebesadel73}, still need further investigation in order to be fully understood  (Kumar \& Zhang, 2005). From a phenomenological point of view, a GRB is composed of the prompt emission, which consists of high-energy photons such as $\gamma$-rays and hard X-rays, and the afterglow emission, i.e. a long-lasting multi-wavelength emission (X-ray, optical, infra-red and sometimes also radio), which follows the prompt. The first afterglow was observed in 1997 by the {\it BeppoSAX} satellite \citep{costa97,vanparadijs1997}. The X-ray afterglow emission has been studied extensively with the BeppoSAX, XMM Newton and especially with Swift (Gerhels et al. 2005) that has discovered the plateau emission (a flat part in the afterglow soon after the decay phase of the prompt emission) and the phenomenology of the early afterglow.

GRBs are traditionally classified into short (with durations $T_{90}<2\,{\rm s}$, SGRBs) and long (with $T_{90}>2\,{\rm s}$, LGRBs) \citep{kouveliotou93}, depending on their duration. Moreover, the possibility of a third class with intermediate durations was argued some time ago \citep{horvath98,mukherjee}, however counter-arguments were recently presented \citep{zitouni,tarnopolski2015a}. Another class---SGRBs with extended emission---exhibiting properties mixed between SGRBs and LGRBs was discovered by \citet{norris2006}. On the other hand, X-ray Flashes (XRFs, \citealt{Heise2001,Kippen2001}), extra-galactic transient X-ray sources with properties similar to LGRBs (spatial distribution, spectral and temporal characteristics), are distinct from GRBs due to having a peak in the $\nu F_{\nu}$ prompt emission spectrum at energies roughly an order of magnitude smaller than the one observed for regular GRBs and by a fluence in the X-ray band ($2-30\,{\rm keV}$) greater than in the $\gamma$-ray band ($30-400\,{\rm keV}$). GRB classifications are important for the investigation of GRB correlations since some of them become more or less evident upon introduction of different GRB classes (for a discussion, see \citealt{Amati2006} and \citealt{Dainotti2010}).

LGRBs were early realized to originate from distant star-forming galaxies, and have been confidently associated with collapse of massive stars related to a supernova (SN) (Galama et al. 1998) \citep{hjorth03,malesani04,sparre11,schulze14}. However, some LGRBs with no clear association with any bright SN have been discovered \citep{fynbo06,dellavalle06}. This implies that there might be other progenitors for LGRBs than core-collapse SNe. Another major uncertainty concerning the progenitors of GRBs is that in the collapsar model \citep{woosley06}, LGRBs are only formed by massive stars with metallicity $Z/Z_\odot$ below $\simeq 0.1-0.3$. On the contrary, a number of GRBs is known to be located in very metal-rich systems \citep{perley15}, and nowadays one of the most important goals is to verify whether there is another way to form LGRBs besides the collapsar scenario \citep{greiner15}. Regarding instead short GRBs, due to their small duration and energy they are consistent with a progenitor of the merger type (NS-Ns, NS-BH). The observations of the location of the short GRBs within their host galaxies tend to confirm this scenario. Additionally, this origin makes them extremely appealing as counterparts of gravitational waves. 

A common model used to explain the GRB phenomenon is the ``fireball'' model (Caballo \& Riess 1974) in which interactions of highly relativistic material within a jet cause the prompt phase, and interaction of the jet with the ambient material leads to the afterglow phase \citep{wijers97,meszaros1998,meszaros2006}. In the domain of the fireball model there are different possible mechanisms of production of the X/Gamma radiation, such as the synchrotron, the Inverse Compton (IC), the Black Body radiation, and sometimes a mixture of these. There are two main flavors of the fireball: the kinetic energy dominated and the magnetic field dominated (Poynting flux dominated). The fireball model encounters difficulties when we would like to match it with observations, because we are not yet able to discriminate neither between the radiation mechanisms nor between these two types of fireball. In addition, we still have a lot to learn about the estimate of the jet opening angles and the structure of the jet itself.
The crisis of the standard fireball model was reinforced when {\it Swift} observations revealed a more complex behavior of the light curves \citep{Obrien06,sakamoto07,Willingale2007,zhang07c} than in the past, hence the discovered correlations among physical parameters are very important in order to discriminate between the fireball and competing theoretical models that have been presented in the literature in order to explain the wide variety of observations. Using these empirical relations corrected for selection biases can allow insight in the GRB emission mechanism. Moreover, given that redshift range over which GRBs can be observed (up to $z=9.4$; \citealt{cucchiara11}) is much larger than given by SNe Ia (up to $z=2.26$; \citealt{Rodney2015}), it is extremely promising to include them as cosmological probes to understand the nature of dark energy (DE) and determine the evolution of the equation of state (EoS), $w=w(z)$, at very high $z$. However, GRBs cannot yet be considered standardizable because of their energies spanning 8 orders of magnitude (see also \citealt{lin2015} and references therein). Therefore, finding universal relations among observable GRB properties can lead to standardization of their energetics and luminosities, which is the reason why the study of GRB correlations is so crucial for understanding the GRB emission mechanism, building a reliable cosmological distance indicator, and estimating the cosmological parameters at high $z$. However, a big caveat needs to be considered in the evaluation of the proper cosmological parameters, since selection biases play a major and crucial role for GRBs, which are particularly affected by the Malmquist bias effect that favors the brightest objects against faint ones at large distances. Therefore, it is necessary to investigate carefully the problem of selection effects and how to overcome it before using GRB correlations as distance estimators, cosmological probes, and model discriminators. This is the main point of this review.

This paper is organized as follows: in Sect.~\ref{notations} we explain the nomenclature and definitions employed throughout this work. In Sect.~\ref{Selection effects} we describe how prompt correlations can be affected by selection biases. In Sect.~\ref{redshiftestimator} we present how to obtain redshift estimators, and in Sect.~\ref{cosmology} we report the use of some correlations as examples of GRB applications as a cosmological tool. Finally, we provide a summary in Sect.~\ref{conclusions}.

\section{Notations and nomenclature}\label{notations}
For clarity and self-completeness we provide a brief summary of the nomenclature adopted in the review. $L$, $F$, $E$, $S$ and $T$ indicate the luminosity, the energy flux, the energy, the fluence and the timescale, respectively, which can be measured in several wavelengths. More specifically:
\begin{itemize}
\item $T_{90}$ is the time interval in which 90\% of the GRB's fluence is accumulated, starting from the time at which 5\% of the total fluence was detected \citep{kouveliotou93}.

\item $T_{\rm peak}$ is the time at which the pulse (i.e., a sharp rise and a slower, smooth decay \citep{fishman94,92,stern96,ryde02}) in the prompt light curve peaks.

\item $T_{\rm break}$ is the time of a power law break in the afterglow light curve \citep{SPH99,willingale2010}, i.e. the time when the afterglow brightness has a power law decline that suddenly steepens due to the slowing down of the jet until the relativistic beaming angle roughly equals the jet opening angle $\theta_{\rm jet}$ \citep{Rhoads97}

\item $\tau_{\rm lag}$ and $\tau_{\rm RT}$ are the difference of arrival times to the observer of the high energy photons and low energy photons defined between $25-50\,{\rm keV}$ and $100-300\,{\rm keV}$ energy band, and the shortest time over which the light curve increases by $50\%$ of the peak flux of the pulse.

\item $T_p$ is the end time prompt phase at which the exponential decay switches to a power law, which is usually followed by a shallow decay called the plateau phase, and $T_a$ is the time at the end of this plateau phase \citep{Willingale2007}.

\item $L_a, L_X$ are the luminosities respective to $T_a,T_p$.

\item $L$ is the observed luminosity, and specifically $L_{\rm peak}$ and $L_{\rm iso}$ are the peak luminosity (i.e., the luminosity at the pulse peak, \citealt{norris2000}) and the total isotropic luminosity, both in a given energy band. 

\item $E_{\rm peak}$, $E_{\rm iso}$, $E_{\gamma}$ and $E_{\rm prompt}$ are the peak energy, i.e. the energy at which 
the $\nu F_{\nu}$ spectrum peaks, the total isotropic energy emitted during the whole burst (e.g., \citealt{AmatiEtal02}), the total energy corrected for the beaming factor (the latter two are connected via $E_\gamma=(1-\cos\theta_{\rm jet})E_{\rm iso}$), and the isotropic energy emitted in the prompt phase, respectively.

\item $F_{\rm peak}$, $F_{\rm tot}$ are the peak and the total fluxes respectively \citep{lee96}.

\item $S_{\gamma}$ and $S_{\rm obs}$ indicate the prompt fluence in the whole gamma band (i.e., from a few hundred keV to a few MeV) and the observed fluence in the range $50-300\,{\rm keV}$.

\item $V$ is the variability of the GRB's light curve. It is computed by taking the difference between the observed light curve and its smoothed version, squaring this difference, summing these squared differences over time intervals, and appropriately normalizing the resulting sum \citep{reichart2001}. Various smoothing filters may be applied (see also \citealt{li2006} for a different approach).
\end{itemize}

Most of the quantities described above are given in the observer frame, except for $E_{\rm iso}$, $E_{\rm prompt}$, $L_{\rm peak}$ and $L_{\rm iso}$, which are already defined in the rest frame. With the upper index ``$*$'' we explicitly denote the observables in the GRB rest frame. The rest frame times are the observed times divided by the cosmic time expansion, for example the rest frame duration is denoted with $T^*_{90}=T_{90}/\left(1+z\right)$. The energetics are transformed differently, e.g. $E^*_{\rm peak}=E_{\rm peak}(1+z)$.

The Band function \citep{Band1993} is a commonly applied phenomenological spectral profile. Its parameters are the low- and high-energy indices $\alpha$ and $\beta$, respectively, and the break energy $E_0$. For the cases $\beta<-2$ and $\alpha>-2$, the $E_{\rm peak}$ can be derived as $E_{\rm peak}=(2+\alpha)E_0$, which corresponds to the energy at the maximum flux in the $\nu F_\nu$ spectra \citep{Band1993,yonetoku04}.

The Pearson correlation coefficient \citep{Kendall,Pearson} is denoted with $r$, the Spearman correlation coefficient \citep{Spearman} with $\rho$, and the $p$-value (a probability that a correlation is drawn by chance) is denoted with $P$. Finally, as most of the relations mentioned herein are power laws, we refer to their slope as to a slope of a corresponding log-log relation.

\section{Selection Effects}\label{Selection effects}

Selection effects are distortions or biases that usually occur when the observational sample is not representative of the 
``true'', underlying population. This kind of bias usually affects GRB correlations. \citet{Efron1992}, \citet{Lloyd99}, 
\citet{Dainotti2013a}, and \citet{petrosian14} emphasized that when dealing with a multivariate data set it is important to focus 
on the intrinsic correlations between the parameters, not on the observed ones, because the latter can be just the result of 
selection effects due to instrumental thresholds. Moreover, how lack of knowledge about the efficiency function influences the 
parameters of the correlations has been already discussed for both the prompt \citep{butler09} and afterglow phases 
\citep{dainotti15,Dainotti2015b}. In this Sect. we revise the selection effects present in the measurements of the GRB prompt parameters 
described in Sect.~\ref{notations}: the peak energy $E_{\rm peak}$ and peak luminosity $L_{\rm peak}$, the isotropic energy 
$E_{\rm iso}$ and isotropic luminosity $L_{\rm iso}$, and the times.

\begin{figure}[htbp]
\centering
\centerline{\includegraphics[width=0.55\columnwidth]{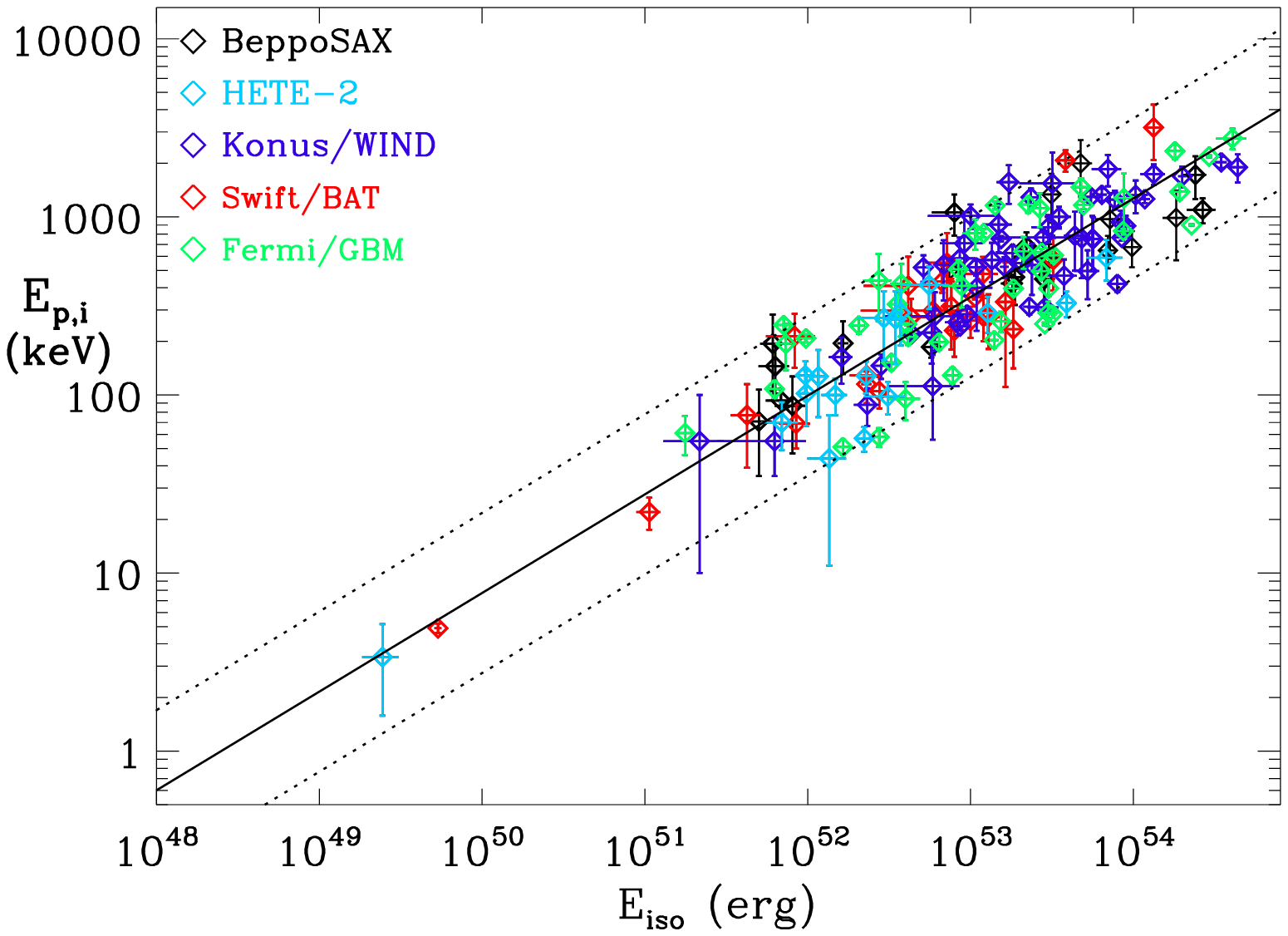}\includegraphics[width=0.62\columnwidth]{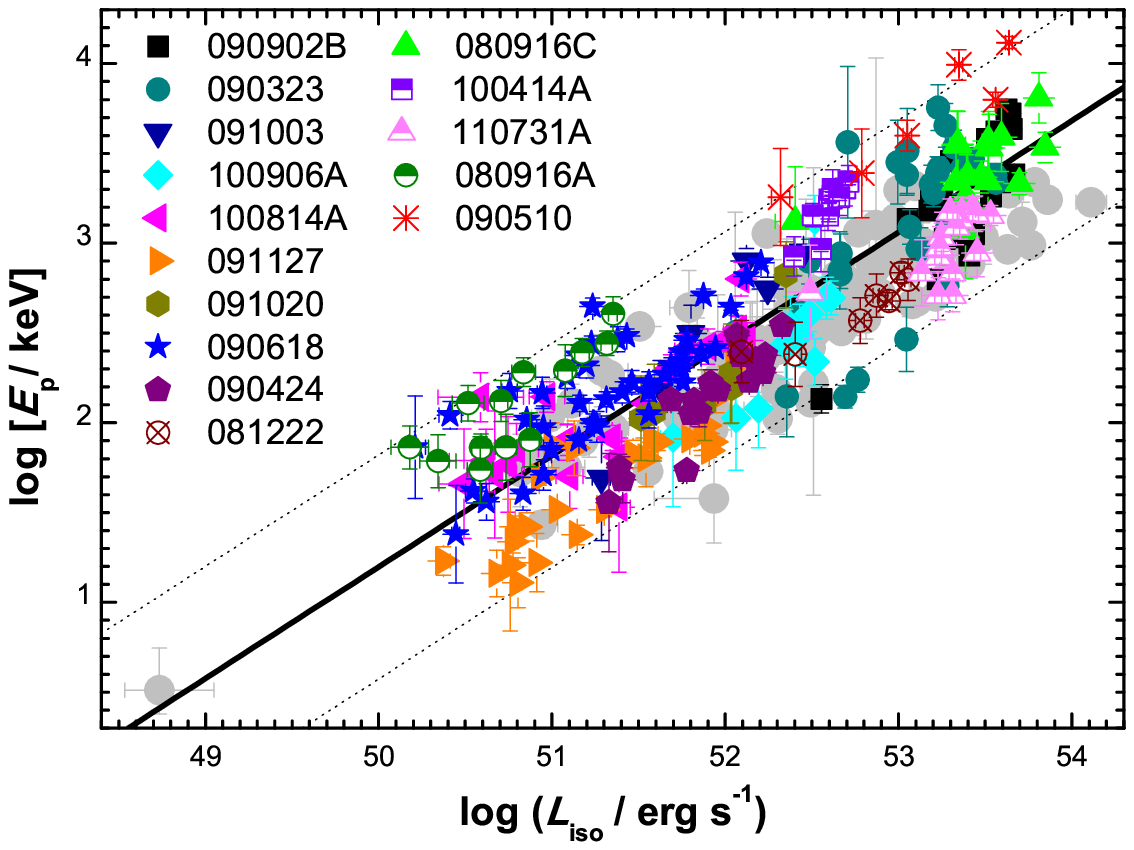}}
\caption{\footnotesize Left: the $E_{\rm peak}$ -- $E_{\rm iso}$ correlation in long GRBs as shown in Amati \& 
Della Valle (2013). Right: the time--resolved $E_{\rm peak}$ -- $L_{\rm iso}$ correlation based on Fermi GRBs as
reported by Lu et al. (2012).}
\label{correlations}
\end{figure}

\begin{figure}[htbp]
\centering
\includegraphics[width=0.7\columnwidth,angle=0]{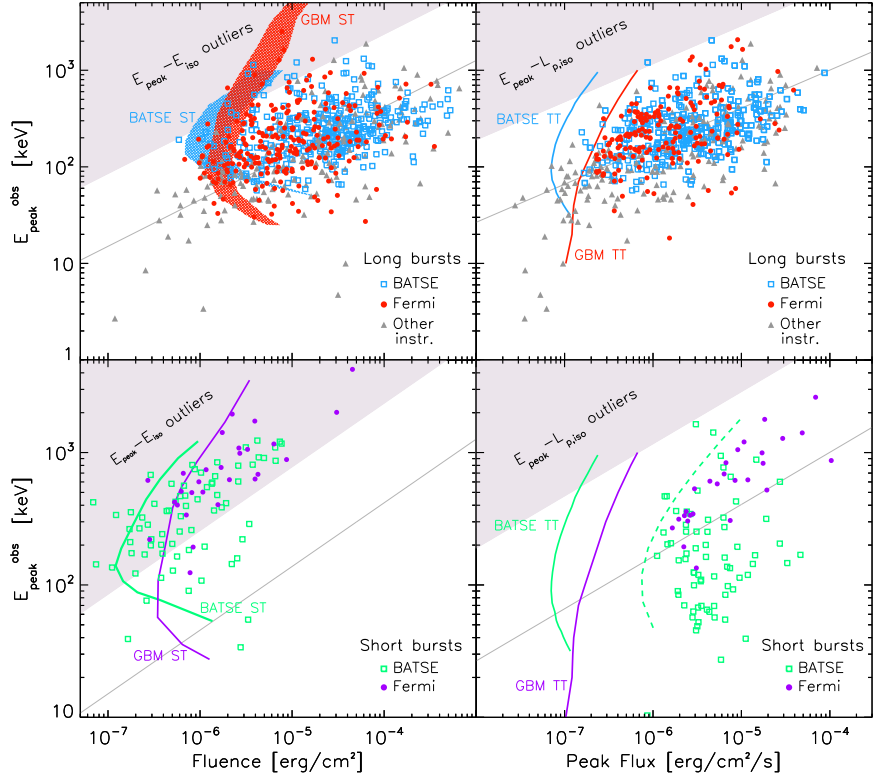}
\caption{\footnotesize 
$E_{\rm peak}$ -- fluence and $E_{\rm peak}$ -- peak flux
planes for long (upper panels) and short (bottom panels) bursts (Nava et al. 2011). Empty squares represent BATSE bursts, filled
circles represent GBM bursts and filled triangles indicate events detected by other instruments. In all panels the instrumental limits for BATSE and
GMB are reported: shaded curved regions in the upper-left panel show the fluence threshold, estimated assuming burst duration of 5 and 20 s; solid curves in the bottom-left
panel represent the fluence threshold for short bursts. Solid curves in the right-hand panels define the trigger threshold, identical for short and long events.
The dashed curve in the bottom-right panel represents the selection
criterion applied, i.e. Peak flux $\ge$ 3 photons cm$^{-2}$ s$^{−1}$. The shaded regions in the upper-left corners of all the planes 
are the region identifying the outliers at more than 3$\sigma$ of the $E_{\rm peak}$ -- $E_{\rm iso}$ (left-hand panels) and $E_{\rm peak}$ -- $L_{\rm peak}$  (right-hand panels) correlations for any given redshift. GRBs, without measured redshift, which fall in these regions are outliers of the corresponding rest-frame correlations ($E_{\rm peak}$ -- $E_{\rm iso}$ and $E_{\rm peak}$ -- $L_{\rm peak}$ for the left- and
right-hand panels, respectively) for any assigned redshift. It means that there is no redshift which makes them consistent with these correlations (considering
their 3$\sigma$ scatter). }
\label{nava11}
\end{figure}

\subsection{Selection effects for the peak energy}\label{selpeak}

\citet{mallozzi95} discussed the photon spectra used for the determination of $E_{\rm peak}$. These parameters were obtained by 
averaging the count rate over the duration of each event. In addition, the temporal evolution of the single light curve affects 
the signal to noise ratio (S/N), i.e. a more spiky light curve will have a larger S/N than a smooth, single-peak event. However, 
\citet{ford95} showed that this effect is not relevant for $E_{\rm peak}$. Considering the most luminous GRBs, they claimed a 
relevant evolution of $E_{\rm peak}$ with time. For this reason, \citet{mallozzi95} used time-averaged spectra, resulting in an 
average value of $E_{\rm peak}$ for each burst. They believed that this evolution should not have a significant impact on their 
results. It should be noted, however, that $E_{\rm peak}$ evolution for bursts with different intensities has not yet been 
examined. Moreover, it has been shown that the fluence, the flux, and the peak energy are affected by data truncation caused by 
the detector threshold \citep{lee96,Lloyd99} and this will generate a bias against high $E_{\rm peak}$ bursts with small fluence 
or flux and an artificial positive correlation in the data. For this reason, it is important to investigate the truncation on the 
data before carrying out research on the correlations.

\citet{lloydpetr2000} focused exemplary on the BATSE detector in order to understand whether truncation effects play a role in 
the $E_{\rm peak}-F_{\rm tot}$ correlation. As discussed in \citep{Lloyd99}, the threshold of any physical parameter determined 
by BATSE is obtained through the trigger condition. For each burst, $C_{\rm max}$, $C_{\rm min}$---the peak photon counts and the 
background in the second brightest detector---are known. Given $S_{\rm obs}$ (or $F_{\rm peak}$), the threshold can be computed 
using the relation

\begin{equation}
\frac{F_{\rm peak}}{F_{\rm peak,lim}} = \frac{S_{\rm obs}}{S_{\rm obs,lim}} = \frac{C_{\rm max}}{C_{\rm min}}.
\label{eq1}
\end{equation}
The condition in Eq.~(\ref{eq1}) is true if the GRB spectrum does not undergo severe spectral evolution. \citet{lloydpetr2000} considered spectral parameters from the Band model. Given a GRB spectrum $f_{\alpha,\beta,A}(E, E_{\rm peak},t)$, the fluence is obtained as
\begin{equation}
S_{\rm obs}=\int\limits_0^T dt \int\limits_{E_1}^{E_2} E\ f_{\alpha,\beta,A}(E, E_{\rm peak},t)dE,
\label{eq2}
\end{equation}
where $T$ is the burst duration. The limiting $E_{\rm peak,lim}$, from which the BATSE instrument is still triggered, is given by the relation
\begin{equation}
S_{\rm obs,lim}=\int\limits_0^T dt \int\limits_{E_1}^{E_2} E\ f_{\alpha,\beta,A}(E, E_{\rm peak,lim},t)dE.
\label{eq3}
\end{equation}

To compute the lower ($E_{\rm peak,min}$) and upper ($E_{\rm peak,max}$) limits on $E_{\rm peak}$, \citet{lloydpetr2000} 
decreased and increased, respectively, the observed value of $E_{\rm peak}$ until the condition in Eq.~(\ref{eq3}) was satisfied. 
In addition, using non-parametric techniques developed by \citet{efron99}, they showed how to correctly remove selection bias 
from observed correlations. This method is general for any kind of correlation. Next, similarly to what was done by 
\citet{Lloyd99}, once $E_{\rm peak,min}$ and $E_{\rm peak,max}$ were determined, \citet{lloydpetr2000} showed that the intrinsic 
$E_{\rm peak}$ distribution is much broader than the observed one. 
Therefore, they analyzed how these biases influence the outcomes. After a careful study of the selection effects, it was 
claimed that an intrinsic correlation between $E_{\rm peak}$ and $E_{\rm iso}$ indeed exists. 
In 
addition, as an important 
constraint on physical models of GRB prompt emission, the $E_{\rm peak}$ distribution is broader than that inferred previously 
from the observed $E_{\rm peak}$ values of bright BATSE GRBs \citep{Zhang02}.

The $E_{\rm peak}$ and $E_{\rm iso}$ correlation, a.k.a. `Amati relation', was actually discovered in 2002 \citep{AmatiEtal02} based on the first sample of BeppoSAX GRBs with measured 
redshift, and later confirmed and extended by measurements by HETE--2, Swift, Fermi/GBM, Konus--WIND 
(Fig.~\ref{correlations}, left panel). The fact that detectors with different sensitivities as a function of photon energy
observe a similar correlation is a first--order indication that instrumental effects should not be dominant.
Soon after, it was shown that the same correlation holds between $E_{\rm peak}$ and
the peak luminosity $L_{\rm peak}$ \citep{yonetoku04}. Moreover, it was pointed out by Ghirlanda et al. (2004) that the $E_{\rm peak}$ and $E_{\rm iso}$ correlation
becomes tighter and steeper (`Ghirlanda relation') when applying the correction for the jet opening angle. This correction however can be applied only for the sub--sample of GRBs from which this quantity could be estimated based on the break observed in the afterglow light curve. In addition, this method of estimating the jet opening angle is model dependent and
affected by several uncertainties.

Later, \citet{Band2005} showed that the Amati and the Ghirlanda relation 
could be converted into a similar energy ratio

\begin{equation}
\frac{E^{1/\eta_i}_{\rm peak}}{S_{\gamma}} \propto F(z).
\end{equation}

Here, $\eta_i$ are the best fit power law indices for the respective correlations. These energy ratios can be represented as 
functions of redshift, $F(z)$, and their upper limits could be determined for any $z$. The upper limit of the energy ratio of both 
the Amati and Ghirlanda relations can be projected onto the peak energy-fluence plane where they become lower limits. In 
this way, it is possible to use GRBs without redshift measurement to test the correlations of the intrinsic peak energy $E_{\rm 
peak}$ with the radiated energy ($E_{\rm iso}$, $E_{\gamma}$) or peak luminosity ($L_{\rm peak}$), as shown
in Fig~\ref{nava11}. By using this method the
above and other authors (Goldstein et al. 2010, Collazzi et al. 2012) found that a significant fraction of BATSE and Fermi GRBs are potentially
inconsistent with the $E_{\rm peak}$ and $E_{\rm iso}$ correlation.

However, several other authors (Bosnjak et al. 2008,Ghirlanda et al. 2008, Nava et al. 2011) showed that, when properly taking into account the dispersion of the correlation
and the uncertainties on spectral parameters and fluences, only a few percent of GRBs may be outliers of the correlation.
Morover, it can be demonstrated (Dichiara et al. 2013) that such a small fraction of outliers can be artificially created
by the combination of instrumental sensitivity and energy band, namely the typical hard--to--soft spectral evolution of
GRBs.

Along this line of investigations, recently, Bosnjak et al. (2014) presented the evaluation of $E_{\rm peak}$ based on the updated {\it INTEGRAL} catalogue of GRBs observed between December 2002 and February 2012. In their spectral analysis they investigated the energy regions with highest sensitivity to compute the spectral peak energies. In order to account for the possible biases in the distribution of the spectral parameters, they compared the GRBs detected by {\it INTEGRAL} with the ones observed by {\it Fermi} and BATSE within the same fluence range. A lower flux limit ($< 8.7\times 10^{-5}\,{\rm erg}\,{\rm cm}^{-2}$ in $50-300\,{\rm keV}$ energy range) was assumed because the peak fluxes from different telescopes were computed in distinct energy ranges. Then, with the proper evaluation of $E_{\rm peak}$, they considered correlations between the following parameters: i) $E_{\rm peak}$ and $F_{\rm tot}$, ii) $E_{\rm peak}$ and $\alpha$, and iii) $E_0$ and 
$\alpha$.

In the case of $E_{\rm peak}-\alpha$ relation no significant correlation was found, while for $E_0-\alpha$ relation there was a weak negative correlation ($\rho=-0.44$) with $P=1.15\times10^{-2}$. In case of the $E_{\rm peak}-F_{\rm tot}$ relation, a weak positive correlation ($\rho=0.50$) was found with $P=1.88\times10^{-2}$. This is in agreement with the results of \citet{kaneko06}, who found a significant correlation between $E_{\rm peak}$ and $F_{\rm tot}$ analyzing the spectra of 350 bright BATSE GRBs with high spectral and temporal resolution.
Regarding the detector-related $E_{\rm peak}$ uncertainties, \citet{collazzi11} noticed that there 
is a discrepancy among the values of $E_{\rm peak}$ found in literature that goes beyond the $1\sigma$ uncertainty.

Finally, notwithstanding the fact that GRBs must be sufficiently bright to perform a time-resolved spectroscopy 
and have known redshifts, 
if the process generating GRBs is independent of the brightness then the existence of the time--resolved 
$E_{\rm peak}-L_{\rm iso}$ 
correlation \citep{Ghirlanda10,Lu12,Frontera12}, see Fig.\ref{correlations} right panel, is a further evidence that these E$_{\rm peak}$ -- `intensity'  
correlations have a physical origin linked to the main emission mechanism in GRBs.

\subsection{Selection effects for the isotropic energy}\label{seliso}
Regarding the selection effects related to $E_{\rm iso}$, \citet{AmatiEtal02} found that the GRBs with measured redshift can be biased due to their paucity, and that the sensitivities and energy bands of the Wide Field Camera (WFC) and Gamma-ray Burst Monitor (GBM) onboard {\it BeppoSAX} and {\it Fermi}, respectively, might prefer energetic and luminous GRBs at larger redshifts, thus creating an artificial $E_{\rm peak}-E_{\rm iso}$ relation.

Therefore, similarly to \citet{Lloyd99} as discussed in Sect.~\ref{selpeak}, \citet{AmatiEtal02} analyzed the $E_{\rm peak,min}$ and $E_{\rm peak,max}$ for which the $E_{\rm peak}-E_{\rm iso}$ correlation exists. If the spectral parameters are coincident with their minimum and maximum values then it is very likely that data truncation will produce a spurious correlation.

Considering a sample of BATSE GRBs without measured redshifts, two research groups (Nakar et al. 2005,Band et al. 2005) claimed that around $50\%$ (Nakar et al. 2004) or even $80\%$ (Band et al. 2005) of  GRBs do not obey the $E_{\rm peak}-E_{\rm iso}$ correlation. This is due to the fact that the selection effects may favor a burst sub-population for which the Amati relation is valid. GRBs with determined redshifts must be relatively bright and soft to be localized. In addition, it was found that selection effects were present in their analyses and that for this reason only the redshifts of the GRBs obeying the relation were computed. However, other authors \citep{ghirlanda05,bosnjak06} arrived at opposite conclusions and these different results are due to considering (or not) the dispersion in the relation, and the uncertainties in $E_{\rm peak}$ and the fluence. Indeed, considering both these features, only some BATSE GRBs with no measured redshift may be considered outliers of the $E_{\rm peak}-E_{\rm iso}$ relation \citep{ghirlanda05}.

Later, \citet{Amati2006} overestimated $E_{\rm peak}$ values because of the paucity of data below $25\,{\rm keV}$. Indeed, if there were selection effects in the sample of {\it HETE-2} GRBs with known redshift, they were more plausible to occur due to detector sensitivity as a function of energy than as a function of the redshift \citep{Amati2006}. The fact that all {\it Swift} GRBs with known redshift are consistent with the $E_{\rm peak}-E_{\rm iso}$ correlation is a strong evidence against the existence of relevant selection effects. \citet{Amati2006} justified this statement adducing the following points: i) the {\it Swift}/BAT sensitivity in $15-30\,{\rm keV}$ is comparable with that of BATSE, {\it BeppoSAX} and {\it HETE-2}, ii) the rapid XRT localization of GRBs decreased the selection effects dependent on the redshift estimate. However, BAT gives an estimate of $E_{\rm peak}$ only for $15-20\%$ of the events. Besides, it was also claimed that the existence of sub-energetic events (like GRB 980425 and possibly GRB 031203) with spectral characteristics are not in agreement with the obtained relation.

\citet{Ghirlanda2008} studied the redshift evolution of the $E_{\rm peak}-E_{\rm iso}$ correlation by binning the GRB sample into different redshift ranges and comparing the slopes in each bin. There is no evidence that this relation evolves with $z$, contrary to what was found with a smaller GRB sample. Their analysis showed, however, that the bursts detected before {\it Swift} are not influenced by the instrumental selection effects, while in the sample of {\it Swift} GRBs the smallest fluence for which it is allowed to compute $E_{\rm peak}$ suffers from truncation effects in 27 out of 76 events.

\citet{amati09} analyzed the scatter of the $E_{\rm peak}-E_{\rm iso}$ relation at high energies and pointed out that it is not influenced by truncation effects because its normalization, computed assuming GRBs with precise $E_{\rm peak}$ from {\it Fermi}/GBM, is in agreement with those calculated from other satellites (e.g., {\it BeppoSAX}, {\it Swift}, {\it KONUS}/Wind). It was also checked whether $E_{\rm iso}$ in the $1\,{\rm keV}-10\,{\rm MeV}$ band can affect the $E_{\rm peak}-E_{\rm iso}$ relation, but its scatter does not seem to vary. Finally, it was also pointed out that: i) the distribution of the new sample of 95 LGRBs is consistent with previous results, ii) in the $E_{\rm peak}-E_{\rm iso}$ plane the scatter is smaller than in the $E_{\rm peak}-F_{\rm tot}$ plane, but if the redshift is randomly distributed then the distributions are similar, and iii) all LGRBs with measured redshift (except GRB 980425) detected with {\it Fermi}/GBM, {\it BeppoSAX}, {\it HETE-2}, and {\it Swift}, obey the $E_{\rm peak}-E_{\rm iso}$ relation \citep{Amati2008}. An exhaustive analysis of instrumental and selection effects for the $E_{\rm peak}-E_{\rm iso}$ correlation is underway and will be reported elsewhere (Dainotti et al. in preparation).

Another example of an analysis of selection effects for $E_{\rm iso}$ was given by \citet{butler09}. They studied the influence of the detector threshold on the $E_{\rm peak}-E_{\rm iso}$ relation, considering a set of 218 {\it Swift} GRBs and 56 {\it HETE-2} ones. Due to the different sensitivities of {\it Swift} and {\it HETE-2} instruments, in the {\it Swift} survey more GRBs are detected. In other words, there is a deficit of data in samples observed in the pre-{\it Swift} missions, and this biases possible correlations. \citet{butler09} tested the reliability of a generic method for dealing with data truncation in the correlations, and afterwards they employed it to data sets obtained by {\it Swift} and pre-{\it Swift} satellites. However, {\it Swift} data does not rigorously satisfy the independence from redshift if there are only bright GRBs, as instead occured for the pre-{\it Swift} $E_{\rm peak}-E_{\rm iso}$ relation. 

Later, Collazzi et al. (2012) argued that the Amati relation may be an artifact of, or at least significantly biased by, a combination of selection effects due to detector sensitivity and energy thresholds. It was found that GRBs following the Amati relation are distributed above a limiting line. Even if bursts with spectroscopic redshifts are consistent with Amati's limit, it is not true for bursts with spectroscopic redshift measured by BATSE and {\it Swift}. In the case in which selection effects are significant, the data in an $E_{\rm peak}-E_{\rm iso}$ plane, obtained by distinct satellites, display different distributions. Eventually, it was pointed out that the selection effects for a detector with a high threshold allow to detect only GRBs in the area where GRBs follow the Amati relation (the so called Amati region), hence these GRBs are not useful cosmological probes.

Instead, the main conclusion drawn from the research of \citet{heussaff13} is that the $E_{\rm peak}-E_{\rm iso}$ relation is generated by a physical constraint that does not allow the existence of high values of $E_{\rm iso}$ and low values of $E_{\rm peak}$, and that the sensitivity of $\gamma$-ray and optical detectors favours GRBs located in the $E_{\rm peak}-E_{\rm iso}$ plane near these constraints. These two effects seem to explain the different results obtained by several authors investigating the $E_{\rm peak}-E_{\rm iso}$ relation.

\citet{amati13}, in order to further discuss the issue of the dependence of the Amati relation on the redshift, analyzed the reliability of the $E_{\rm peak}-E_{\rm iso}$ relation using a sample of 156 GRBs available until the end of 2012. They divided this sample into subsets with different redshift ranges (e.g., $0.1 < z < 1$, $1 < z < 2$, etc.), pointing out that the selection effects are not significant because the slope, normalization and scatter of the correlation remain constant. They found
\begin{equation}
\log \frac{E_{\rm peak}}{1\,{\rm keV}} = 0.5 \log \frac{E_{\rm iso}}{10^{52}\,{\rm erg}}+2.
\end{equation}

Finally, \citet{mochkovitch14}, with a model that took into account the small amount of GRBs with large $E_{\rm iso}$ and small $E_{\rm peak}$, pointed out that the scatter of the intrinsic $E_{\rm peak}-E_{\rm iso}$ relation is larger than the scatter of the observed one.

\subsection{Selection effects for the isotropic luminosity}\label{sellum}
\citet{ghirlanda2012b} studied a data set of 46 GRBs and claimed that the flux limit---introduced to take into account selection biases related to $L_{\rm iso}$---generates a constraint in the $L_{\rm iso}-E_{\rm peak}$ plane. Given that this constraint corresponds to the observed relation, they pointed out that $87\%$ of the simulations gave a statistically meaningful correlation, but only $12\%$ returned the slope, normalization and scatter compatible with those of the original data set. There is a non-negligible chance that a boundary with asymetric scatter may exist due to some intrinsic features of GRBs, but to validate this hypothesis additional complex simulations would be required.

\begin{figure}
\includegraphics[width=12cm, height=8.5cm,angle=0]{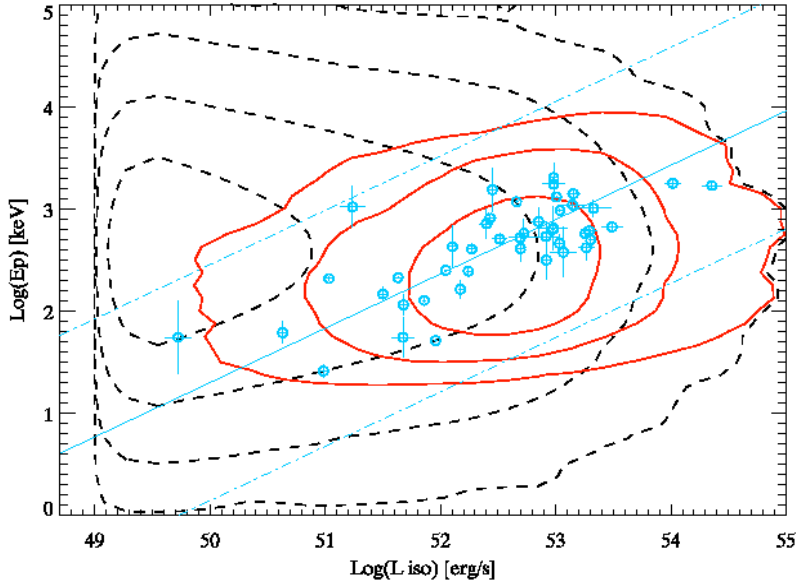}
\caption{\footnotesize Simulation showing the variation of the density from \citep{ghirlanda2012b}. The simulated sample is represented by the dashed contour (1, 2, 3 and $4\sigma$ levels). The sample of simulated GRBs with a flux greater than the constraint $F_{\rm lim}=2.6\,{\rm ph}\,{\rm cm}^{-2}\,{\rm s}^{-1}$ in the $15-150\,{\rm keV}$ energy band is depicted by red solid contours (1, 2 and $3\sigma$ levels). The blue circles denote the 46 {\it Swift} GRBs used by \citet{nava12}: the solid line represents the best fit of the $L_{\rm iso}-E_{\rm peak}$ relation, and the dot-dashed lines display the $3\sigma$ region around the best fit line.}
\label{fig:ghirlanda2012}
\end{figure}

Additionally, they performed Monte Carlo simulations of the GRB population under different assumptions for their luminosity functions. Assuming there is no correlation between $E_{\rm peak}$ and $L_{\rm iso}$, they were unable 
to reproduce it, thus confirming the existence of an intrinsic correlation between $E_{\rm peak}$ and $L_{\rm iso}$ at more than $2.7\sigma$. For this reason, there should be a relation between these two parameters that does not originate from detector limits (see Fig~\ref{fig:ghirlanda2012}).
	
\subsection{Selection effects for the peak luminosity}\label{sellupeak}

\citet{Yonetoku2010} investigated how the truncation effects and the redshift evolution affect the $L_{\rm peak}-E_{\rm peak}$ relation. They claimed that the selection bias due to truncations might occur when the detected signal is comparable with the detector threshold, and showed that the relation is indeed redshift dependent.

Shamoradi et al. (2013) studied the $E_{\rm peak}-E_{\rm iso}$ and $L_{\rm peak}-E_{\rm peak}$ relations, and constructed a model describing both the luminosity function and the distribution of the prompt spectral and temporal parameters, taking into account the detection threshold of $\gamma$-ray instruments, particularly BATSE and {\it Fermi}/GBM. Analyzing the prompt emission data of 2130 BATSE GRBs, he demonstrated that SGRBs and LGRBs are similar in a 4-dimensional space of $L_{\rm peak}$, $E_{\rm iso}$, $E_{\rm peak}$ and $T_{90}$. Moreover, he showed that these two relations are strongly biased by selection effects, questioning their usefulness as cosmological probes. Similar $E_{\rm peak}-E_{\rm iso}$ and $L_{\rm peak}-E_{\rm peak}$ relations, with analogous correlation coefficient and significance, should hold for SGRBs. Based on the multivariate log-normal distribution used to model the luminosity function it was predicted that the strong correlation between $T_{90}$ and both $E_{\rm iso}$ and $L_{\rm peak}$ was valid for SGRBs as well as for LGRBs.

\citet{shahmoradi14} investigated the luminosity function, energetics, relations among GRB prompt parameters, and methodology for classifying SGRBs and LGRBs using 1931 BATSE events. Employing again the multivariate log-normal distribution model, they found out statistically meaningful $L_{\rm peak}-E_{\rm peak}$ and $E_{\rm peak}-E_{\rm iso}$ relations with $\rho=0.51\pm 0.10$ and $\rho=0.60\pm 0.06$, respectively.

\citet{yonetoku04}, \citet{Petrosian2015} and \citet{Dainotti2015b} showed how $L_{\rm peak}$ undergoes redshift evolution. These authors found a strong redshift evolution, $L_{\rm peak}\propto (1+z)^{2.0-2.3}$, an evolution which is roughly compatible in $1.5\sigma$ among the authors. Different data samples were used, but statistical method, i.e. the \citet{Efron1992} one, was the same. This tool uses a non-parametric approach, a modification of the Kendall $\tau$ statistics. A simple [$f(z)=(1+z)^\alpha$] or a more complex redshift function was chosen and in both cases compatible results were found for the evolution. After a proper correction of the $L_{\rm peak}$ evolution, \citet{Dainotti2015b} established the intrinsic $L_{\rm peak}-L_X$ relation. This statistical method is very general and it can be applied also to properties of the afterglow emission. For example, \citet{Dainotti2013a} found the evolution functions for the luminosity at the end of the plateau emission, $L_X$, and its rest frame duration, $T^*_a$, a.k.a the Dainotti relation, and they demonstrated the intrinsic nature of the $L_X-T^*_a$ relation. From the existence of the intrinsic nature of $L_X-T^*_a$ and $L_{\rm peak}-L_X$ relations Dainotti et al. (2016a) discovered the extended $L_X-T^*_a-L_{\rm peak}$ relation, which is intrinsic as being a combination of two intrinsic correlations. However, in this paper the authors had to deal with additional selection bias problems. They analyzed two samples, the full set of 122 GRBs and the gold sample composed of 40 GRBs with high quality data in the plateau emission. The selection criteria were defined carefully enough not to introduce any biases, which was shown to be indeed valid by a statistical comparison of the whole and gold samples. They found a tight $L_a-T^*_a-L_{\rm peak}$ relation, and showed that for the gold sample it has a 54\% smaller scatter than a corresponding $L_a-T^{*}_a$ relation obtained for the whole sample. Moreover, it was shown via the bootstrapping method that the reduction of scatter is not due to the smaller sample size, hence the 3D relation for the gold sample is intrinsic in nature, and is not biased by the selection effects of this sample. From this paper follows that a thorough consideration of the selection effects can provide insight into the physics underlying GRB emission. In fact, this fundamental plane relation can be a very reliable test for physical models. An open question would be if the magnetar model \citep{zhang2001,troja07,rowlinson14,rea15} can still be a plausible explanation of this relation as it was for the $L_X-T^*_a$ relation. Thus, once correlations are corrected for selection bias can be very good candidates to explore, test and possibly efficiently discriminate among plausible theoretical models.

\subsection{Selection effects for the lag time and the rise time}\label{seltime}
Considering $\tau_{\rm lag}$ and its dependence on the redshift, \citet{Azzam2012} studied the evolutionary effects of the $L_{\rm peak}-\tau_{\rm lag}$ relation using 19 GRBs detected by {\it Swift}, and found the results to be in perfect agreement with those obtained through other methods (like the ones from \citealt{tsutsui2008}, who used redshifts obtained through the Yonetoku relation to study the dependence of the $L_{\rm peak}-\tau_{\rm lag}$ one on redshift). Specifically, he divided the data sample in 3 redshift ranges (i.e., $0.540-1.091$, $1.101-1.727$, $1.949-3.913$) and calculated the slope and normalization of the log-log relation in each of them. In the first bin a slope of $-0.92\pm0.19$ and a normalization of $51.94\pm0.11$ were found with $r=-0.89$. In the second bin the slope was $-0.82\pm 0.12$ and the normalization $52.12\pm0.08$ with $r=-0.94$. In the third bin the relation had a slope equal to $-0.04 \pm 0.22$ and normalization $52.90\pm0.12$ with $r=-0.06$. Therefore, the $L_{\rm peak}-\tau_{\rm lag}$ relation seems to evolve with redshift, however, this conclusion is only tentative since there is the problem that each redshift range is not equipopulated, and it is limited by low statistics and by the significant scatter in the relation. Therefore, the $L_{\rm peak}-\tau_{\rm lag}$ relation is redshift-dependent, but this result is not conclusive due to the paucity of the sample and a significant scatter.

\citet{kocevski2013} found that in individual pulses the observer-frame cosmological time dilation is masked out because only the most luminous part of the light curve can be observed by GRB detectors. Therefore, the duration and $E_{\rm iso}$ for GRBs close to the detector threshold need to be considered as lower limits, and the temporal characteristics are not sufficient to discriminate between LGRBs and SGRBs (see also \citealt{tarnopolski2015b} for a novel and successful attempt of using for this purpose non-standard parameters via machine learning).

Regarding instead the rise time, \citet{Wang11} used 72 LGRBs observed by {\it BeppoSAX} and {\it Swift} to study the $L_{\rm iso}-\tau_{\rm RT}$ relation, and found that the relation is not dependent on the redshift. In fact, for the total sample they obtained
\begin{equation}
\log \frac{L_{\rm iso}}{1 {\rm erg}\,{\rm s}^{-1}} = (52.68 \pm 0.07)-(1.12 \pm 0.14)\log \frac{\tau^*_{\rm RT}}{0.1\,{\rm s}},
\end{equation}
with $\sigma_{\rm int}=0.48 \pm 0.05$. Additionally, dividing the data set into 4 redshift bins (i.e., $0-1$, $1-2$, $2-3$, and $3-8.5$), the slope and normalization of this relation remained nearly constant. This represents good evidence for the $L_{\rm iso}-\tau_{\rm RT}$ relation not being influenced by evolutionary effects.

\section{Redshift Estimators}\label{redshiftestimator}
As we have already pointed out, it is relevant to study GRBs as possible distance estimators, since for many of them $z$ is unknown. Therefore, having a relation which is able to infer the distance from observed quantities independent on $z$ would allow a better investigation of the GRB population. Moreover, in the cases in which $z$ is uncertain, its estimators can give hints on the upper and lower limits of the GRB's distance. Some examples of redshift estimators from the prompt relations \citep{atteia03,yonetoku04,tsutsui13} have been reported.

In \citep{atteia03}, the $E_{\rm peak}-E_{\rm iso}$ relation, due to the dependence of these quantities on $D_L(z,\Omega_M, \Omega_{\Lambda})$, was analyzed to derive pseudo-redshifts of 17 GRBs detected by {\it BeppoSAX}. They were obtained in the following way: in a first step a combination of physical parameters was considered: $X=\frac{n_{\gamma}\sqrt{T_{90}}}{E_{\rm peak}}$, where $n_{\gamma}$ is the observed number of photons, and then the theoretical evolution of this parameter with $z$ was computed according to
\begin{equation}
X=A f(z),
\end{equation}
where $A$ is a constant and $f(z)$ is the redshift evolution for the energy spectra of a ``standard'' GRB. A ``standard'' GRB has $\alpha=-1.0$, $\beta=-2.3$, and $E_0 = 250\,{\rm keV}$ in a $\Lambda$CDM universe ($H_0 =65\,{\rm km}\,{\rm s}^{-1}\,{\rm Mpc}^{-1}$, $\Omega_M=0.3$, $\Omega_{\Lambda}= 0.7$). The relation representing the redshift evolution was inverted to derive a redshift estimator from the observed quantities given by the equation $z=\frac{1}{A}f^{-1}(X)$. The possible applications of these redshift estimators include a statistically-driven method to compare the distance distributions of different GRB populations, a rapid identification of far away GRBs with redshifts exceeding three, and estimates of the high-$z$ star formation rate.

\citet{yonetoku04}, using 689 bright BATSE LGRBs, analyzed the spectra of GRBs adopting the Band model. The lower flux limit was set to $F_{\rm lim}= 2\times10^{-7}\,{\rm  erg}\,{\rm cm}^{-2}\,{\rm s}^{-1}$ for achieving a higher S/N ratio. Next, $F_{\rm peak}$ and $E^*_{\rm peak}$ were obtained and the pseudo-redshifts of GRBs in the sample were estimated inverting with respect to $z$ the following equation [taking into account the redshift dependence on $D_L(z,\Omega_M,\Omega_{\Lambda})$]:
\begin{equation}
\log L_{\rm peak}=(47.37\pm0.37)+(2.0\pm0.2)\log E^*_{\rm peak}.
\end{equation}

\begin{figure}[htbp]
\centering
\includegraphics[width=9.1cm,height=6.5cm]{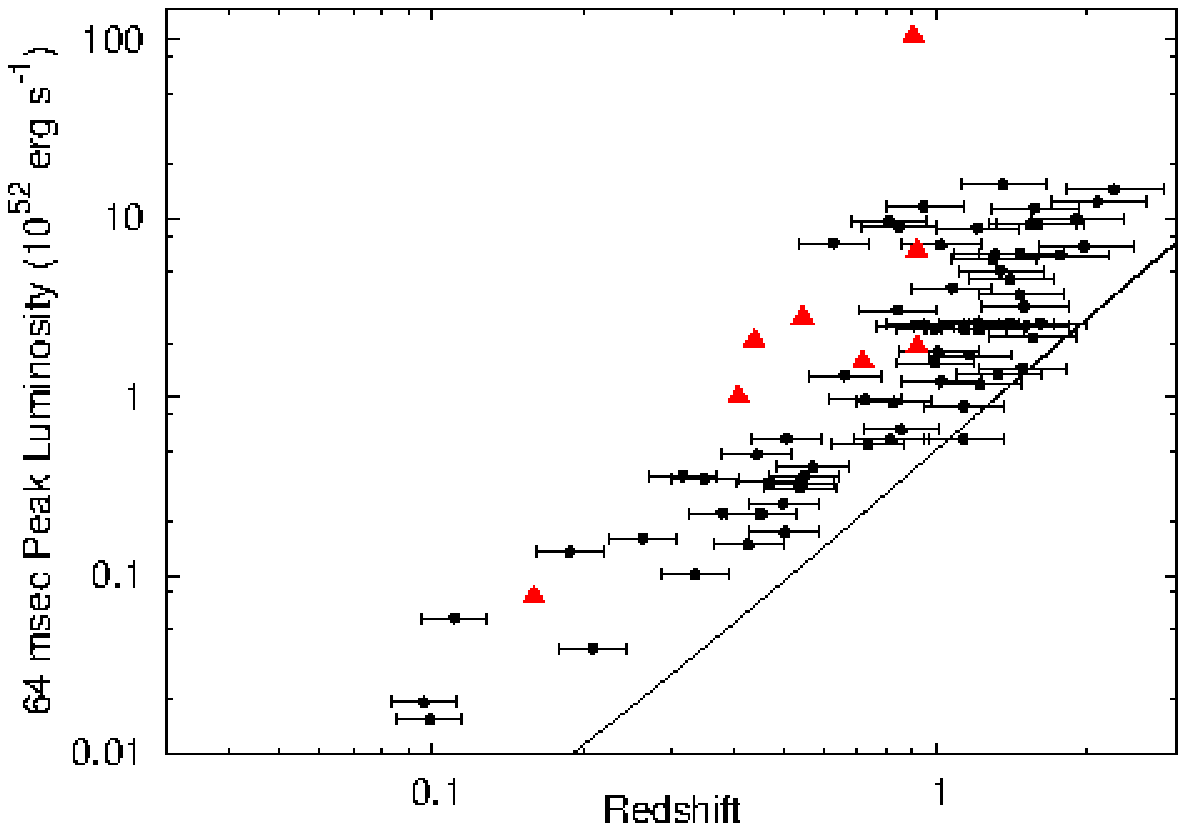}
\caption{\footnotesize The distribution of $z$ estimated by the best fit $\log L_{\rm peak}-\log E_{\rm peak}$ relation for SGRBs from \citep{tsutsui13}; 71 bright BATSE SGRBs from \citep{ghirlanda09} were used. Black dots denote the pseudo-redshifts related to $L_{\rm peak}$, and red filled triangles mark secure SGRBs; $z\in(0.097,2.581)$ with $\langle z\rangle=1.05$, compared with {\it Swift} LGRBs with $\langle z\rangle\approx 2.16$. The solid line represents the flux limit of $F_{\rm peak} = 10^{-6}\,{\rm erg}\,{\rm cm}^{-2}\,{\rm s}^{-1}$.}
\label{fig:tsutsui13} 
\end{figure}

Later, in \citep{tsutsui13} the $E_{\rm peak}-E_{\rm iso}$ and the $L_{\rm peak}-E_{\rm peak}$ relations were tested using a sample of 71 SGRBs detected by BATSE. Comparing these two relations, it was claimed that the $L_{\rm peak}-E_{\rm peak}$ one would be a better redshift estimator because it is tighter. Therefore, rewriting the $L_{\rm peak}-E_{\rm peak}$ relation in the following way:
\begin{equation}
\frac{D^2_L(z, \Omega_M, \Omega_{\Lambda})}{(1 + z)^{1.59}}=\frac{10^{52.29}}{4\pi F_{\rm peak}}\left(\frac{E_{\rm peak}}{774.5}\right)^{1.59},
\end{equation}
and assuming a cosmological model with ($\Omega_M, \Omega_{\Lambda}$) = (0.3, 0.7), the pseudo-redshifts were calculated from $D_L(z,\Omega_M, \Omega_{\Lambda})$, which is a function of $z$ (see Fig. \ref{fig:tsutsui13}). \citet{dainotti11a} investigated the possibility of using the $L_X-T^{*}_a$ relation as a redshift estimator. It appeared that only if one is able to reduce the scatter of the relation to 20\% and be able to select only GRBs with low error bars in their variables, satisfactory results can be achieved. More encouraging results for the redshift estimator are expected with the use of the new 3D (Dainotti et al. 2016a) correlation due to its reduced scatter.

\section{Cosmology}\label{cosmology}
The Hubble Diagram (HD) of SNe Ia, i.e the distribution of the distance modulus\footnote{The difference between the apparent magnitude m, ideally corrected from the effects of interstellar absorption, and the absolute magnitude M of an astronomical object.} $\mu(z)$, opened the way to the investigation of the nature of DE. As is well known, $\mu(z)$ scales linearly with the logarithm of the luminosity distance $D_L(z, \Omega_M, \Omega_{\Lambda})$ (which depends on the DE EoS through a double integration) as follows:
\begin{equation}
\mu(z)  =  25 + 5\log D_L(z, \Omega_M, \Omega_{\Lambda}).
\end{equation}
Discriminating among different models requires extending the HD to higher redshifts since the expression for $\mu(z)$ is different as one goes to higher $z$ values.

\subsection{The problem of the calibration}

One of the most important issues presented in the use of GRB correlations for cosmological studies is the so called circularity problem. Since local GRBs, i.e. GRBs with $z \leq 0.01$, are not available, an exception being GRB 980425 with $z=0.0085$ \citep{galama98}, one has to typically assume an a priori cosmological model to compute $D_L(z, \Omega_M, \Omega_{\Lambda})$ so that the calibration of the two dimensional (2D) correlations turns out to be model dependent. In principle, such a problem could be avoided in three ways. First, through the calibration of these correlations by several low-$z$ GRBs (in fact, at $z\leq 0.1$ the luminosity distance has a negligible dependence on the choice of cosmological parameters). Second, a solid theoretical model has to be found in order to physically motivate the observed 2D correlations, thus setting their calibration parameters. Particularly this would fix their slopes independently of cosmology, but this task still has to be achieved.

Another option for calibrating GRBs as standard candles is to perform the fitting using GRBs with $z$ in a narrow range, $\Delta z$, around some representative redshift, $z_c$. We here describe some examples on how to overcome the problem of circularity using prompt correlations. GRB luminosity indicators were generally written in the form of $L=a\prod x^{b_i}_i$, where $a$ is the normalization, $x_i$ is the $i$-th observable, and $b_i$ is its corresponding power law index. However, \citet{liang06} employed a new GRB luminosity indicator, $E_{\rm iso} = aE^{*b_1}_{\rm peak}T^{*b_2}_{\rm peak}$ (the LZ relation, \citealt{liang05}), and showed that while the dependence of $a$ on the cosmological parameters is strong, it is weak for $b_1$ and $b_2$ as long as $\Delta z$ is sufficiently small. The selection of $\Delta z$ is based on the size and the observational uncertainty of a particular sample. \citet{liang05} proposed to perform the calibration on GRBs with $z$ clustered around an intermediate $z_c\in (1,2.5)$, because most GRBs are observed with such redshifts. Eventually, it was found that $\sim 25$ GRBs around $z_c=1$ with $\Delta z=0.3$ is sufficient for the calibration of the LZ relation to serve as a distance indicator.

Also \citet{ghirlanda06}, using the $E_{\rm peak}-E_{\gamma}$ correlation \citep{Ghirlanda2004} defined $E_{\rm peak}= aE^b_{\gamma}$ as a luminosity indicator. Considering a sample of 19 GRBs detected by {\it BeppoSAX} and {\it Swift}, the minimum number of GRBs required for calibration of the correlation, $N$, was estimated within a range $\Delta z$ centered around a certain $z_c$. For a set of $\Omega_M$ and $\Omega_{\Lambda}$ the correlation was fitted using a sample of $N$ GRBs simulated in the interval $(z_c-\Delta z,z_c+\Delta z)$. The relation was considered to be calibrated if the change of the exponent $b$ was smaller than $1\%$. The free parameters of this test are $N$, $\Delta z$ and $z_c$. Different values of $z_c$ and $\Delta z \in (0.05, 0.5)$ were tested by means of a Monte Carlo technique. At any $z$ the smaller the $N$, the larger was the variation $\Delta b$ of the exponent (for the same $\Delta z$) due to the correlation being less constrained. Instead, for larger $z_c$, smaller $\Delta z$ was sufficient for maintaining $\Delta b$ small. It was found that only 12 GRBs within $z \in (0.9, 1.1)$ are enough to calibrate the $E_{\rm peak}-E_{\gamma}$ relation.

Unfortunately, this method might not be working properly due to the paucity of the observed GRBs. Another method to perform the calibration in a model-independent way might be developed using SNe Ia as a distance indicator based on a trivial observation that a GRB and an SN Ia with the same redshift $z$ must have the same distance modulus $\mu(z)$. In this way GRBs should be considered complementary to SNe Ia at very high $z$, thus allowing to extend the distance ladder to much larger distances. Therefore, interpolation of the SNe Ia HD provides an estimate of $\mu(z)$ for a subset of the GRB sample with $z \leq 1.4$, which can then be used for calibration \citep{kodama2008,liang08,wei09}. 
The modulus is given by the formula \citep{cardone09}
\begin{eqnarray}
\mu(z) & = & 25 + 5\log D_L(z, \Omega_M, \Omega_{\Lambda}) \nonumber \\
       & = & 25 + (5/2)(\log y - k) \\
       & = & 25 + (5/2) (a + b\log x-k) \nonumber 
\label{modulus}
\end{eqnarray}
where $\log y = a+b\log x$, $y = k D_L^2(z, \Omega_M, \Omega_{\Lambda})$ is a given quantity with $k$ a redshift independent constant, $x$ is a distance-independent property, and $a$ and $b$ are the correlation parameters. With the asumption that this calibration is redshift-independent, the HD at higher $z$ can be built with the calibrated correlations for the the GRBs at $z>1.4$ still present in the sample.

\subsection{Applications of GRB prompt correlations}

\citet{dai04} and \citet{xu05} proposed a method to constrain cosmological parameters using GRBs and applied it to preliminary samples (consisting of 12 and 17 events, respectively) relying on the $E_{\rm peak}-E_{\gamma}$ relation. They found $\Omega_M=0.35^{+0.15}_{-0.15}$ \citep{dai04} and $\Omega_M=0.15^{+0.45}_{-0.13}$ \citep{xu05} on $1\sigma$ CL, consistent with SNe Ia data.

\begin{figure}[htbp]
\includegraphics[width=5.4cm,height=5.3cm]{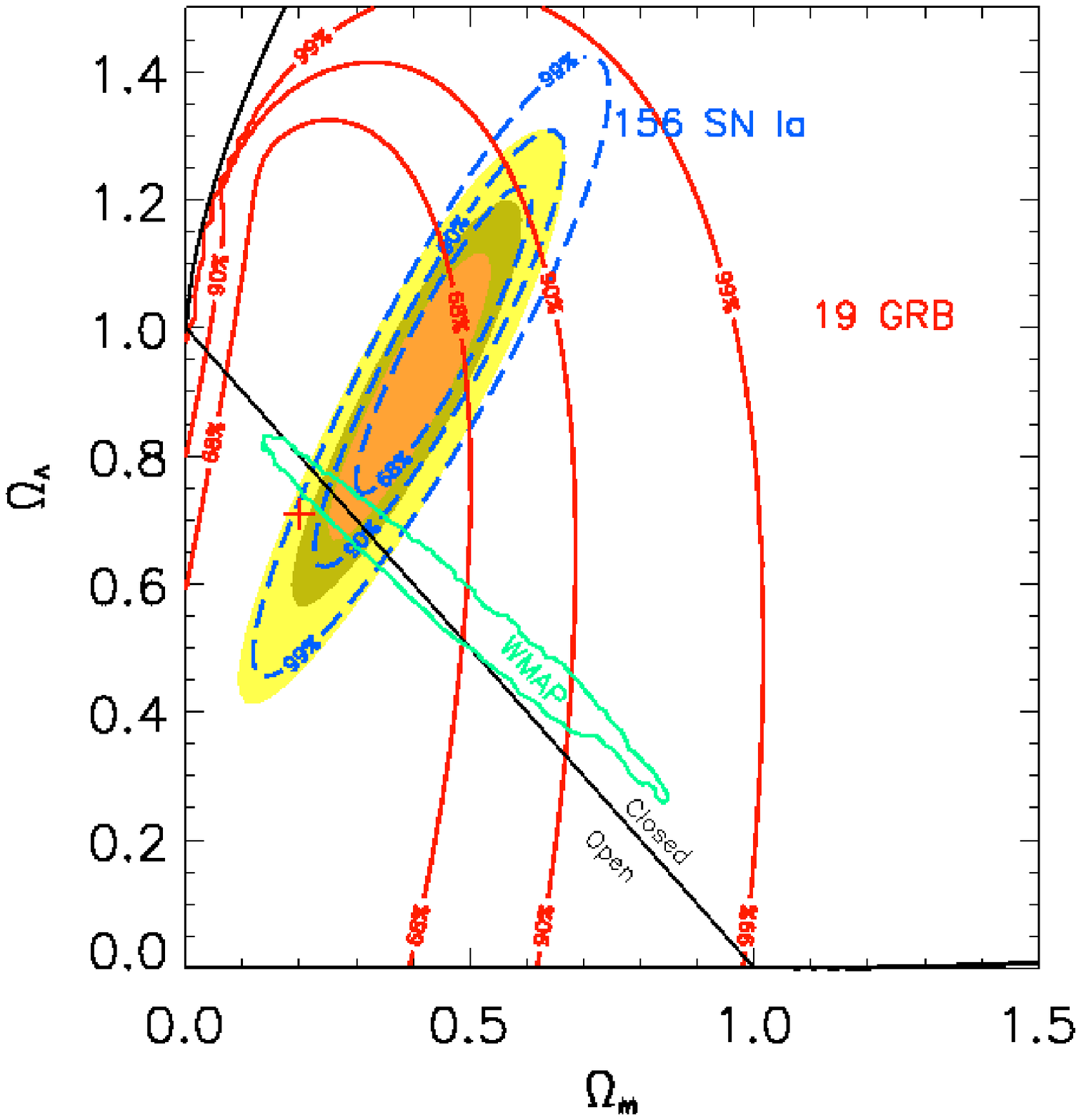}
\includegraphics[width=5.4cm,height=5.3cm]{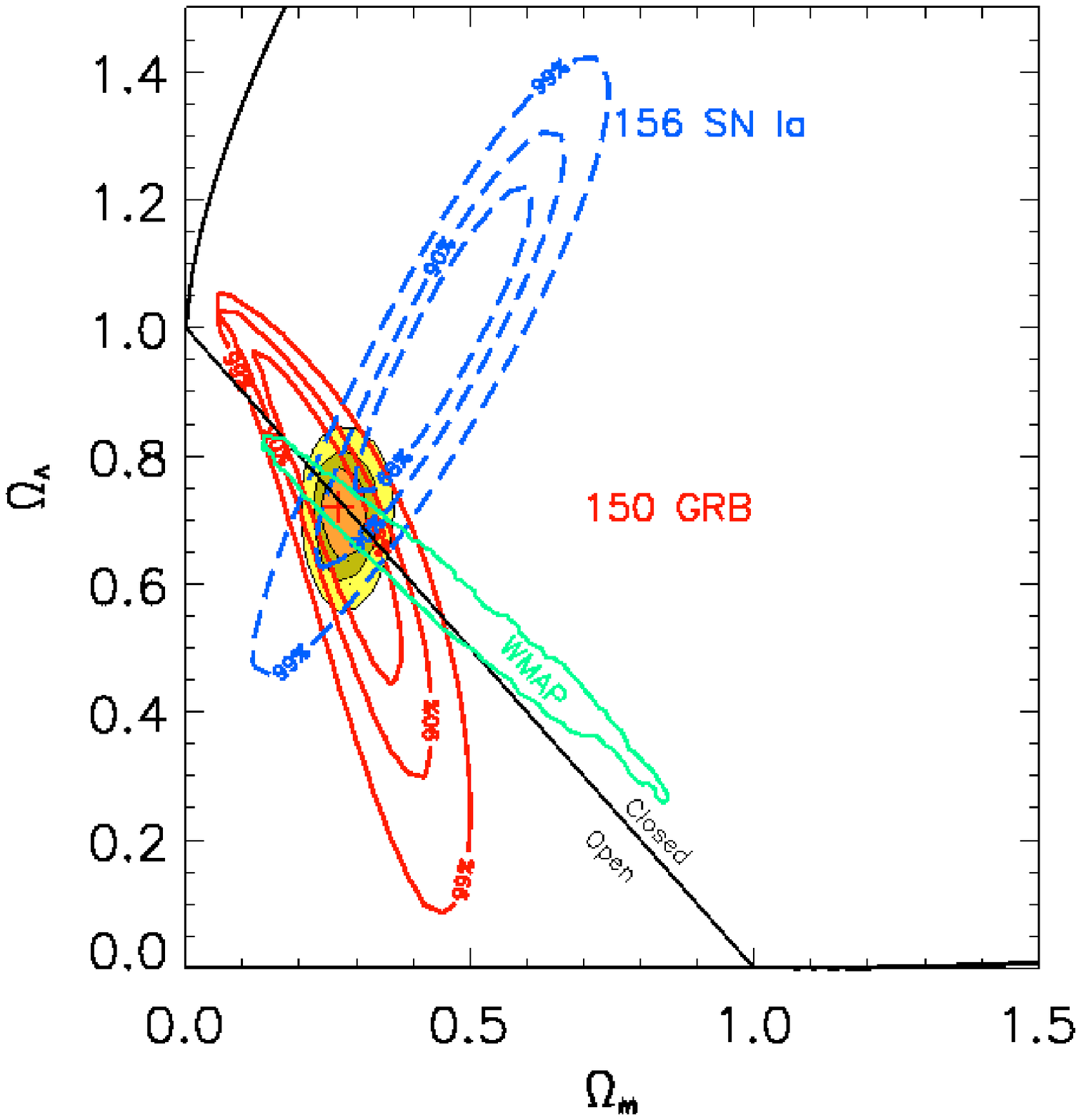}
\includegraphics[width=5.4cm,height=5.3cm]{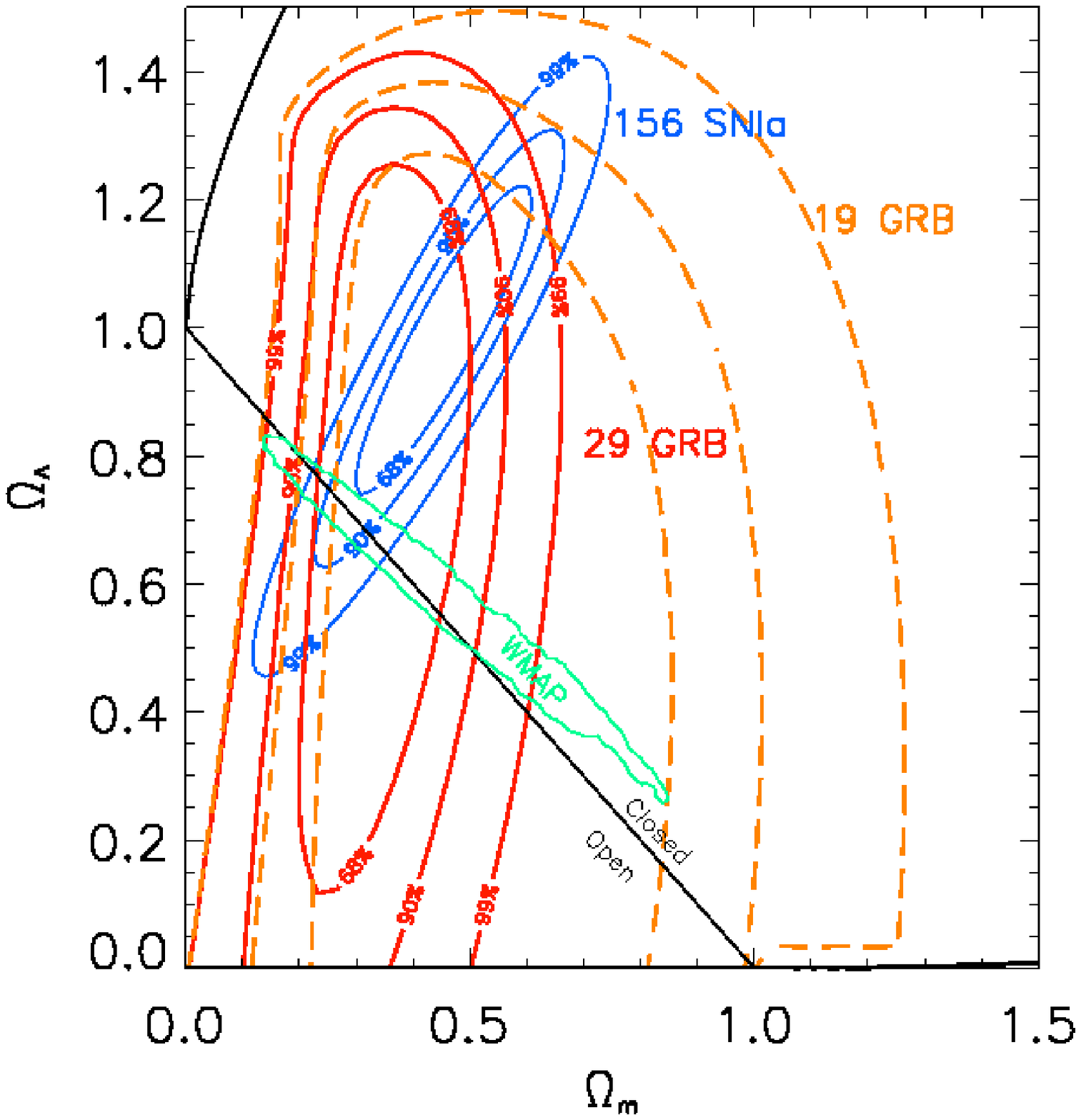}
\caption{\footnotesize {\bf Left panel:} The solid red contours, obtained with the samples of 19 alone, represent the 68.3\%, 90\% and 99\% confidence regions of $\Omega_{M}$ and $\Omega_{\Lambda}$ obtained using the $E_{\rm peak}-E_{\gamma}$ in the HM case \citep{ghirlanda06}. The center of these contours (red cross) corresponds to a minimum $\chi^2$ and yields $\Omega_{M}=0.23$, $\Omega_{\Lambda}=0.81$.
{\bf Middle panel:} The same but obtained with the sample of 150 GRBs simulated by assuming the $E_{\rm peak}-E_{\gamma}$ relation derived in the WM case \citep{ghirlanda06}. On both panels, the contours obtained with 156 SNe Ia of the ``Gold'' sample from \citep{riess04} are shown by the dashed blue lines. The joint GRB+SN constraints are represented by the shaded contours. The 90\% confidence contours obtained with the WMAP data are also shown.
{\bf Right panel:} The same but obtained with the $E_{\rm peak}-E_{\gamma}$ relation updated until January 2009 (29 bursts -- solid line; \citealt{ghirlanda09b}) compared to the previous data (19 GRBs -- dashed line). The constraints obtained from 156 SN type Ia (blue thin line - \cite{riess04}) and those from the WMAP data (green thin line) are also shown.}
\label{fig:ghirlanda}
\end{figure}

Later, \citet{ghirlanda06} used 19 GRBs and claimed that the $E_{\rm peak}-E_{\gamma}$ and $E_{\rm peak}-E_{\rm prompt}-T_{\rm break}$ relations can be used to constrain the cosmological parameters in both the homogeneous (HM, see left panel in Fig.~\ref{fig:ghirlanda}) and wind circumburst medium (WM, see middle panel in Fig.~\ref{fig:ghirlanda}) cases. An updated sample of 29 GRBs \citep{ghirlanda09b} supported previous results (see right panel in Fig. \ref{fig:ghirlanda}).

However, in order to overcome the circularity problem affecting these correlations, \citet{ghirlanda06} considered three different methods to fit the cosmological parameters through GRBs:
\begin{enumerate}[(I)]
\item The scatter method based on fitting the correlation for a set of cosmological parameters that need to be constrained (e.g., $\Omega_{M},\Omega_{\Lambda}$). To fulfill this task, a $\chi^2$ surface in dependence on these parameters is built. The minimum of the $\chi^{2}$ surface indicates the best cosmological model.
\item The luminosity distance method consisting of the following main steps: (1) choose a cosmology (i.e., its parameters) and fit the $E_{\rm peak}-E_{\gamma}$ relation; (2) estimate the term $\log E_{\gamma}$ from the best fit; (3) from the definition of $\log E_{\gamma}$, derive $\log E_{\rm iso}$, from which $D_L(z,\Omega_M, \Omega_{\Lambda})$ is next computed; (4) evaluate $\chi^2$ by comparing $D_L(z,\Omega_M, \Omega_{\Lambda})$ with the one derived from the cosmological model. After iterating these steps over a set of cosmological parameters, a $\chi^{2}$ surface is built. In this case the best cosmology is represented also by the minimum $\chi^2$.
\item The Bayesian method: methods (I) and (II) stem from the concept that some correlation, e.g. $E_{\rm peak}-E_{\gamma}$, exists between two variables. However, these methods do not exploit the fact that the correlations are also very likely to be related to the physics of GRBs and, therefore, they should be unique. Firmani et al. (2005,2006), employing the combined use of GRBs and SNe Ia, proposed a more complex method which considers both the existence and uniqueness of the correlation.
\end{enumerate}

Another correlation that turns out to be useful in measuring the cosmological parameter $\Omega_M$ is the $E_{\rm peak}-E_{\rm iso}$ one \citep{Amati2008}. Adoption of the maximum likelihood approach  provides a correct quantification of the extrinsic scatter of the correlation, and gives $\Omega_M$ narrowed (for a flat universe) to the interval $0.04-0.40$ (68\% CL), with a best fit value of $\Omega_M = 0.15$, and $\Omega_M=1$ is excluded at $>99.9\%$ CL. No specific assumptions about the $E_{\rm peak}-E_{\rm iso}$ relation are made, and no other calibrators are used for setting the normalization, therefore the problem of circularity does not affect the outcomes and the results do not depend on the ones derived from SNe Ia. The uncertainties in $\Omega_M$ and $\Omega_{\Lambda}$ can be greatly reduced, based on predictions of the current and expected GRB experiments.

An example of Bayesian method that takes into account the SNe Ia calibration is presented in \citet{cardone09}. They updated the sample used in \cite{schaefer2007} adding to the previous 5 correlations the Dainotti relation $L_a-T^{*}_a$ \citep{Dainotti2008}. They used a Bayesian-based method for fitting and calibrating the GRB correlations assuming a representative $\Lambda$CDM model. To avoid the problem of circularity, local regression technique was applied to calculate $\mu(z)$ from the most recent SNe Ia sample containing (after selection cuts and having outliers removed) 307 SNe with $0.015 \leq z \leq 1.55$. Only GRBs within the same range of $z$ defined by the SNe data were considered for calibration. It was shown that the estimated $\mu(z)$ for each GRB common to the \citep{schaefer2007} and \citep{Dainotti2008} samples is in agreement with that obtained using the set of \citet{schaefer2007} correlations. Hence, no additional bias is introduced by adding the $L_a-T^{*}_a$ correlation. In fact, its use causes the errors of $\mu(z)$ to diminish significantly (by $\sim14$\%) and the sample size to increase from $69$ to $83$ GRBs.

\begin{figure}[htbp]
\centering
\includegraphics[width=8.1cm,height=6.1cm]{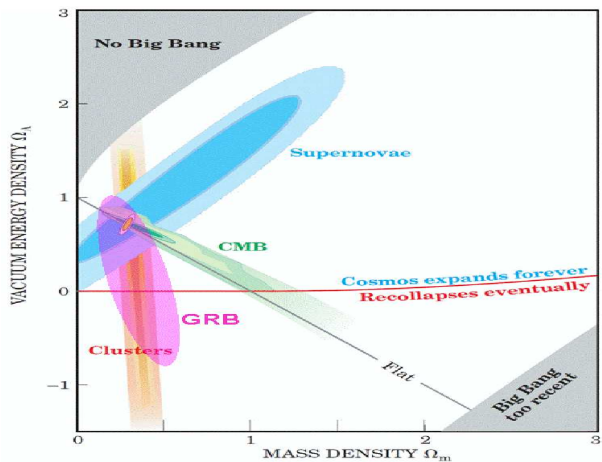}
\caption{\footnotesize 68\% CL contours in the $\Omega_M-\Omega_{\Lambda}$ plane from \citep{amati13} obtained by employing a simulated sample of 250 GRBs expected in the near future (pink) compared to those from other cosmological probes (adapted from a figure by the Supernova Cosmology Project).}
\label{fig:amati13}
\end{figure}

\citet{amati12} and \citet{amati13} used an enriched sample of 120 GRBs to update the analysis of \citet{Amati2008}. Aiming at the investigation of the properties of DE, the 68\% CL contours in the $\Omega_M-\Omega_{\Lambda}$ plane obtained by using a simulated sample of 250 GRBs (pink area in Fig.~\ref{fig:amati13}) were compared to those from other cosmological probes such as SNe Ia, CMB and galaxy clusters (blue, green and yellow areas, respectively, in Fig.~\ref{fig:amati13}).
To obtain the simulated data set, Monte Carlo technique was employed. It took into account the observed $z$ distribution of GRBs, the exponent, normalization and dispersion of the observed $E_{\rm peak}-E_{\rm iso}$ power law relation, and the distribution of the uncertainties in the measured values of $E_{\rm peak}$ and $E_{\rm iso}$. The simulation showed that a sample of $\approx 250$ GRBs is sufficient for the accuracy of $\Omega_M$ to be comparable to the one currently provided by SNe Ia. In addition, estimates of $\Omega_M$ and $w_0$ expected from the current and future observations were given. The authors assumed that the calibration of the $E_{\rm peak}-E_{\rm iso}$ relation is done with an accuracy of 10\% using, e.g., the $D_L$ provided by SNe Ia and the self-calibration of the GRB correlation via a sufficiently large number of GRBs within a narrow range of $z$ (e.g. $\Delta z \sim 0.1-0.2$). It was also noted that as the number of GRBs in each redshift bin is increased, also the accuracy and plausibility of the $E_{\rm peak}-E_{\rm iso}$ relation's self-calibration should increase.

\begin{figure}[htbp]
\includegraphics[width=5.4cm,height=5.3cm]{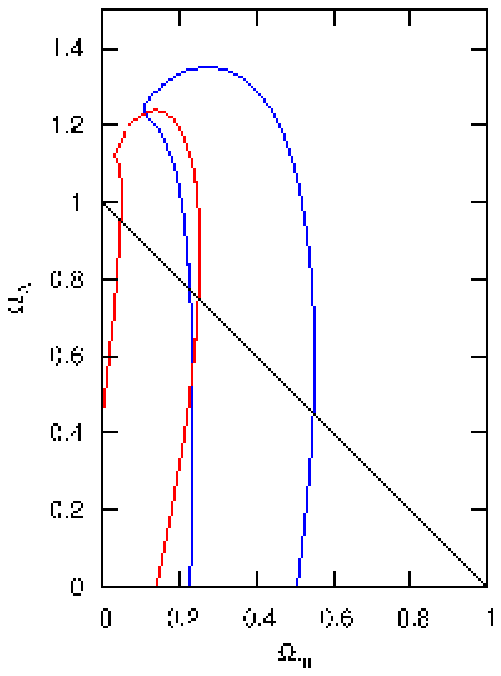}
\includegraphics[width=5.4cm,height=5.1cm]{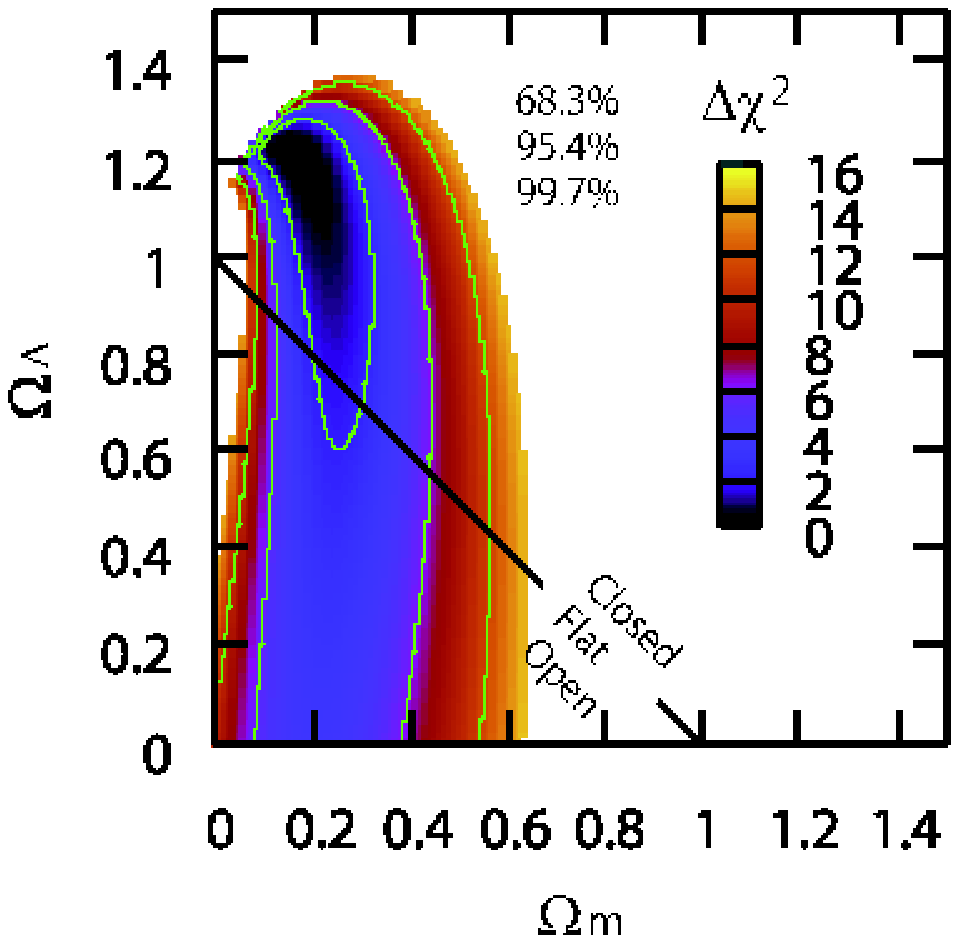}
\includegraphics[width=5.4cm,height=5.3cm]{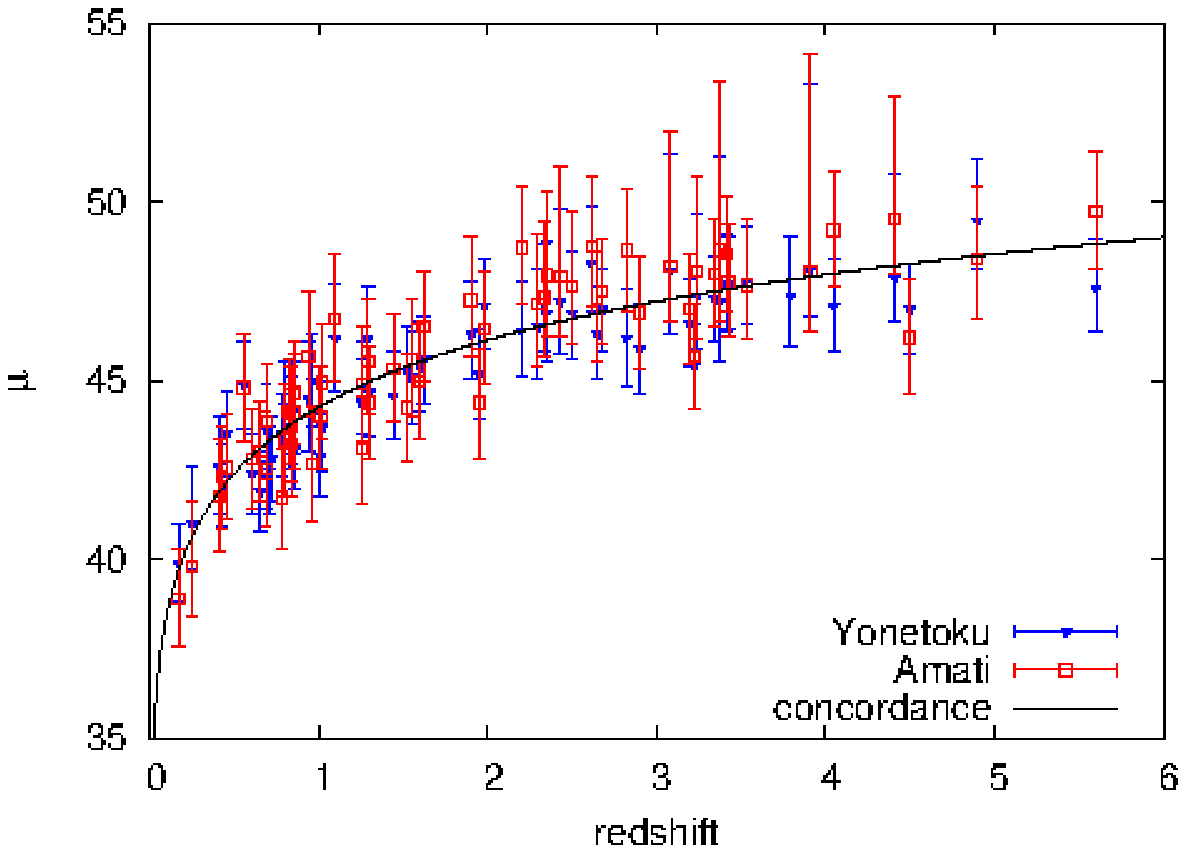}
\caption{\footnotesize {\bf Left panel:} Constraints on $\Omega_M$, $\Omega_{\Lambda}$ from the Amati (red) and Yonetoku (blue) relations \citep{Tsutsui2009}. The contours correspond to 68.3\% confidence regions; black solid line represents the flat universe. The results are consistent on the $2\sigma$ level.
{\bf Middle panel:} Constraints on $\Omega_M$, $\Omega_{\Lambda}$ the from $E_{\rm peak}-T_L-L_{\rm peak}$ relation \citep{Tsutsui2009}.
{\bf Right panel:} Extended HD from the Amati (red) and Yonetoku (blue) relations \citep{Tsutsui2009}. A systematic difference is apparent in high-$z$ GRBs, although it does not seem to be present in low-$z$ ones.}
\label{fig:tsutsui09}
\end{figure}

\citet{Tsutsui2009} employed a sample of 31 low-$z$ GRBs and 29 high-$z$ GRBs to compare the constraints imposed on cosmological parameters ($\Omega_M$ and $\Omega_{\Lambda}$) by i) the $E_{\rm peak}-T_L-L_{\rm peak}$ relation, and ii) the $L_{\rm peak}-E_{\rm peak}$ and $E_{\rm peak}-E_{\rm iso}$ relations calibrated with low-$z$ GRBs (with $z \leq 1.8$; see left and middle panels in Fig.~\ref{fig:tsutsui09}, respectively). Assuming a $\Lambda$CDM model with $\Omega_k = \Omega_M+\Omega_{\Lambda}-1$, where $\Omega_k$ is the spatial curvature density, it was found using the likelihood method that the constraints for the Amati and Yonetoku relations are different in $1\sigma$, although they are still consistent in $2\sigma$. Therefore, a luminosity time parameter, $T_L =E_{\rm iso}/L_{\rm peak}$, was introduced in order to correct the large dispersion of the $L_{\rm peak}-E_{\rm peak}$ relation. A new relation was given as
\begin{equation}
\log L_{\rm peak}=(49.87\pm 0.19)+(1.82\pm 0.08)\log E_{\rm peak}-(0.34\pm 0.09)\log T_L.
\label{eq4}
\end{equation}
The systematic error was successfully reduced by 40\%. Finally, application of this new relation to high-$z$ GRBs (i.e., with $1.8<z<5.6$) yielded $(\Omega_M, \Omega_{\Lambda}) = (0.17^{+0.15}_{-0.08}, 1.21^{+0.07}_{-0.61})$, consistent with the $\Lambda$CDM model (see right panel in Fig.~\ref{fig:tsutsui09}).

\begin{figure}[htbp]
\includegraphics[width=5.4cm,height=5.5cm]{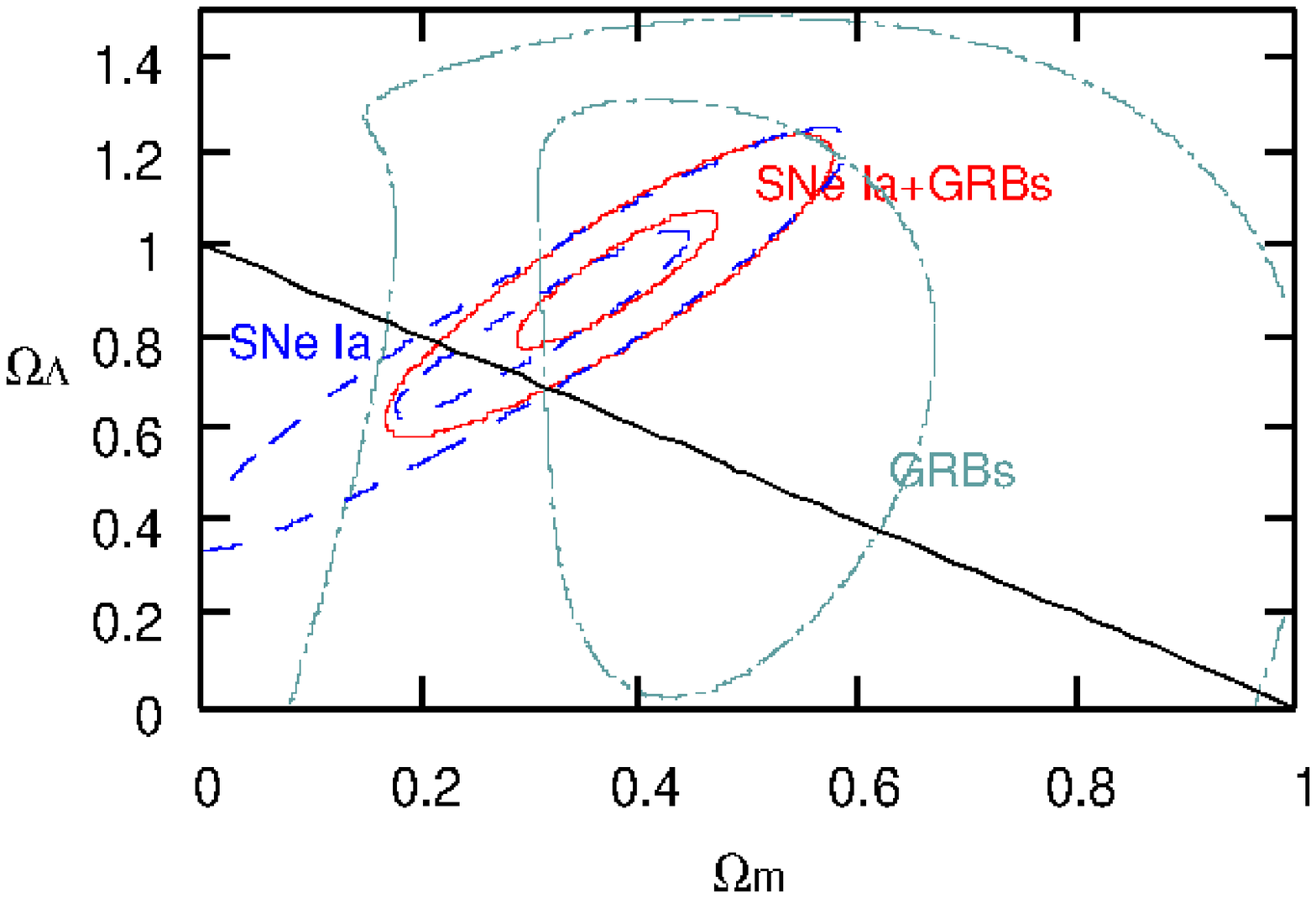}
\includegraphics[width=5.4cm,height=5.8cm]{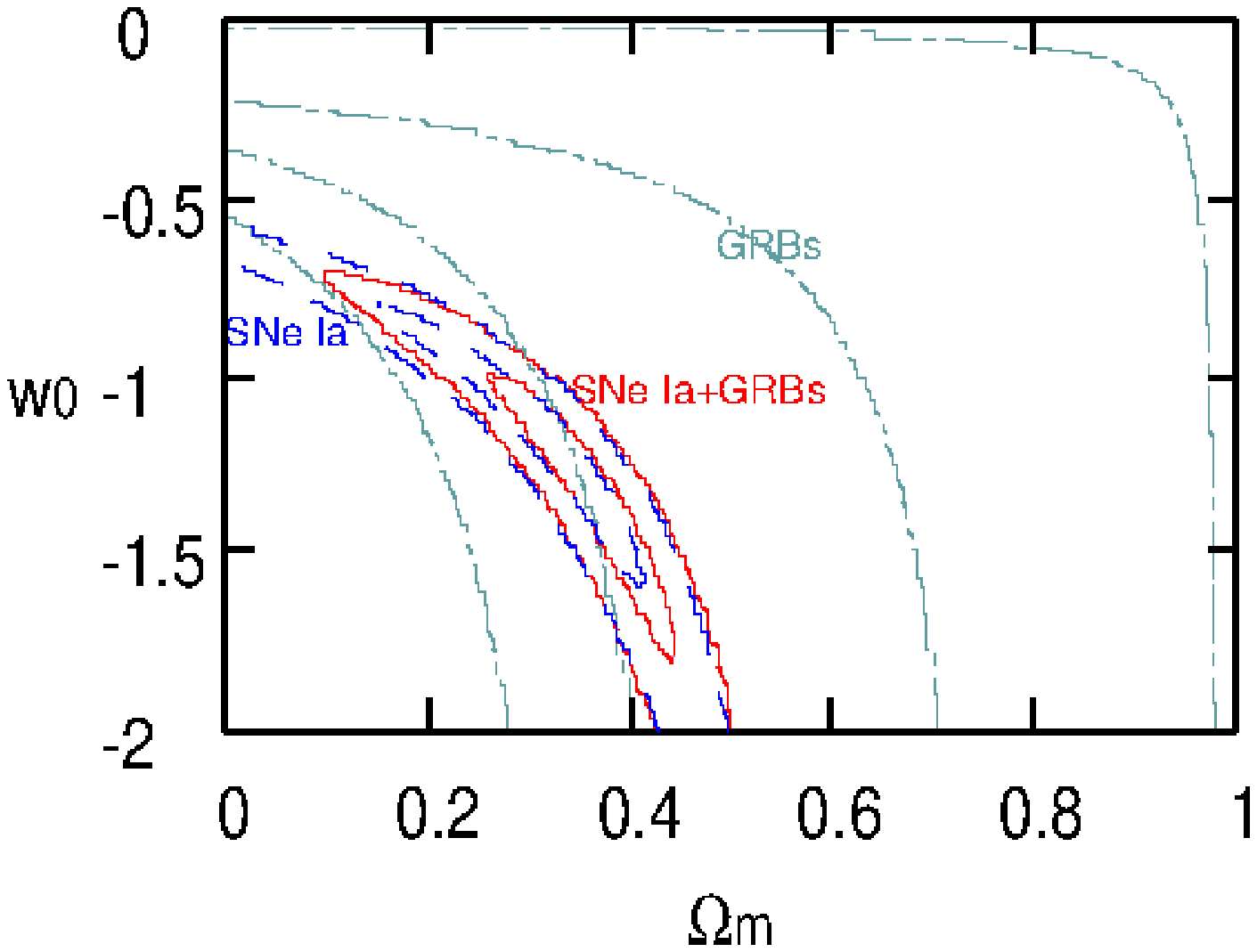}
\includegraphics[width=5.4cm,height=6.1cm]{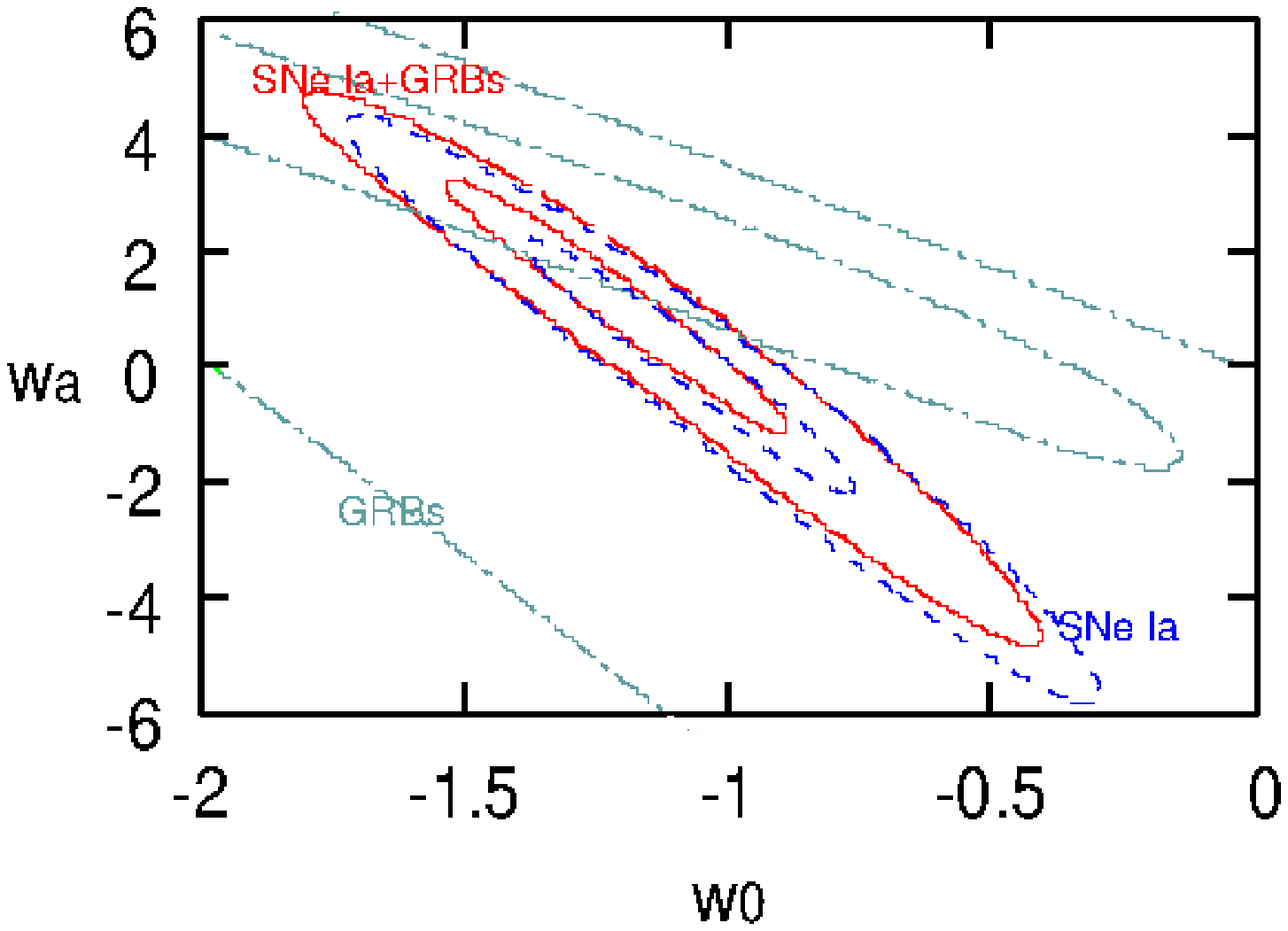}
\caption{\footnotesize The contours of likelihood $\Delta\chi^2$ in ({\bf left panel}) the $(\Omega_M, \Omega_{\Lambda})$, ({\bf middle panel}) the $(\Omega_M$, $w_0)$, and ({\bf right panel}) the 0$(w_0$, $w_a)$ planes for GRBs (light blue dash-doted lines), SNe Ia (blue dotted lines), SNe Ia+GRBs (red solid lines), respectively, from \citep{tsutsui08b}. The contours correspond to 68.3\% and 99.7\% confidence regions, and black solid line represents the flat universe.}
\label{fig:tsutsui08}
\end{figure}

\citet{tsutsui08b}, using the $L_{\rm peak}-E_{\rm peak}$ relation, considered three cosmological cases: a $\Lambda$CDM model, a non-dynamical DE model ($w_a=0$), and a dynamical DE model, viz. with $w(z)=w_0+w_az/(1+z)$, in order to extend the HD up to $z=5.6$ with a sample of 63 GRBs and 192 SNe Ia (see Fig.~\ref{fig:tsutsui08}). It was found that the current GRB data are in agreement with the $\Lambda$CDM model (i.e., $\Omega_M = 0.28$, $\Omega_{\Lambda} = 0.72$, $w_0 = -1$, $w_a = 0$) within $2\sigma$ CL. Next, the constraints on the DE EoS parameters expected from {\it Fermi} and {\it Swift} observations were modeled via Monte Carlo simulations, and it was claimed that the results should improve significantly with additional 150 GRBs.

As shown by \citet{Tsutsui2010}, with the use of the $E_{\rm peak}-T_L-L_{\rm peak}$ relation it is possible to determine the $D_L$ with an error of about 16\%, which might prove to be useful in unveiling the nature of DE at $z>3$. In addition, it was pointed out in \citep{Wang11} that the correlations between the transition times of the X-ray light curve from exponential to power law, and the X-ray luminosities at the transitions such as the \citet{Dainotti2008} and the \citet{Qi2010} correlations may be used as standard candles after proper calibration. This procedure, as explained by \citet{Wang11}, consists of a minimization of $\chi^2$ (with the maximum likelihood method) over the parameters of the log-log relation (i.e., the slope and normalization), and simultaneously over the cosmological parameters. These relations can allow insight into cosmic expansion at high $z$, and at the same time they have the potential to narrow the constraints on cosmic expansion at low $z$. GRBs could also probe the cosmological parameters in order to distinguish between DE and modified gravity models \citep{Wang2009a,Wang2009b,Vitagliano2010,Capozziello2008}.

\citet{lin15} tested the possible redshift dependence of several correlations ($L_{\rm peak}-\tau_{\rm lag}$, $L_{\rm peak}-V$, $E_{\rm peak}-L_{\rm peak}$, $E_{\rm peak}-E_{\gamma}$, $L_{\rm iso}-\tau_{\rm RT}$, and $E_{\rm peak}-E_{\rm iso}$) by splitting 116 GRBs (with $z\in[0.17,8.2]$ from \citealt{Wang11}) into low-$z$ (with $z<1.4$) and high-$z$ (with $z>1.4$) groups. It was demonstrated that the $E_{\rm peak}-E_{\gamma}$ relation for low-$z$ GRBs is in agreement with that for high-$z$ GRBs within $1\sigma$ CL. The scatter of the $L_{\rm peak}-V$ relation was too large to formulate a reliable conclusion. For the remaining correlations, it turned out that low-$z$ GRBs differ from high-$z$ GRBs at more than $3\sigma$ CL. Hence, the $E_{\rm peak}-E_{\gamma}$ relation was chosen to calibrate the GRBs via a model-independent approach. High-$z$ GRBs give $\Omega_M=0.302\pm 0.142$ ($1\sigma$ CL) for the $\Lambda$CDM model, fully consistent with the Planck 2015 results (planck et al. 2015). In conclusion, GRBs have already provided a direct and independent measurement of $\Omega_M$, and simulations show that they will be able to achieve an accuracy comparable to SNe Ia.

\section{Summary}\label{conclusions}
In this work we have reviewed the characteristics of empirical relations among the GRB prompt phase observables, with particular focus on the selection effects, and discussed possible applications of several correlations as distance indicators and cosmological probes. It is crucial that a number of the correlations face the problem of double truncation which affects, e.g., the value of $E_{\rm peak}$. Some relations have also been shown to be intrinsic in nature (e.g. the $E_{\rm peak}-F_{\rm tot}$, $E_{\rm peak}-E_{\rm iso}$, or $L_{\rm peak}-E_{\rm peak}$ ones), while the intrinsic forms are not known for others. As a consequence, we are not yet aware whether these correlations can affect the evaluation of the theoretical models of GRBs and the cosmological setting (Dainotti et al. 2013b, 2016a). Therefore, establishing the intrinsic correlations is crucial. In fact, though there are several theoretical interpretations describing each correlation, in many cases more then one is viable, thus showing that the emission processes that rule GRBs still have to be further investigated. To this end, it is necessary to use the intrinsic correlations, not the observed ones that are affected by selection biases, to test the theoretical models. A very challenging future step would be to use the correlations corrected for biases to determine a further and more precise estimate of the cosmological parameters.

\acknowledgments 
MGD acknowledges the Marie Curie Program due to funding from the European Union Seventh Framework Program (FP7-2007/2013) under grant agreement N 626267. M.G.D. is particular grateful to Lorenzo Amati for his comments and remarks on introduction and sec. 3.1 and 3.2 and to Roberta del Vecchio and Mariusz Tarnopolski for their preliminary work on this review. 

\addcontentsline{toc}{section}{References}
\bibliography{biblioReview5}

\end{document}